# Descriptor-Enabled Rational Design of High-Entropy Materials Over Vast Chemical Spaces

Dibyendu Dey[1,*], Liangbo Liang[2], and Liping Yu[1,*]
[1]Department of Physics and Astronomy, University of Maine, Orono, Maine 04469, USA.
[2]Center for Nanophase Materials Sciences, Oak Ridge National Laboratory, Oak Ridge, Tennessee 37831, USA.
[*]Correspondences: dibyendu.dey@maine.edu, liping.yu@maine.edu

## Abstract

The practically unlimited high-dimensional composition space of high-entropy materials (HEMs) has emerged as an exciting platform for functional materials design and discovery. However, the identification of stable and synthesizable HEMs and robust design rules remains a daunting challenge due to the difficulty in determining composition/structure-specific enthalpy and entropy contributions to the stability and formation of HEMs. In this work, using first-principles calculations, we find that (i) the stability and miscibility of HEMs strongly depend on the formation enthalpy of the HEM relative to the "*most stable*" completing phase rather than on the conventionally-viewed enthalpies of mixing over "*all possible*" competing phases, and (ii) the entropy forming ability of a HEM can be measured from the defect-formation energy spectrum of tens of substitutional defects in ordered binary compounds, involving no sampling over numerous alloy configurations. Based on these findings, we propose a highly predictive Mixed Enthalpy-Entropy Descriptor (MEED), which enables the rational high-throughput first-principles design and screening of new HEMs over large chemical spaces. Applying the MEED to two structurally distinct material systems (i.e., 3D rocksalt carbides and 2D layered sulfides), not only all experimentally reported HEMs within each system are successfully identified, but a universal cutoff criterion for assessing their relative synthesizability also revealed. In addition, tens of new high entropy carbides and 2D high-entropy sulfides are also predicted. They have the potential for a wide variety of applications such as coating in aerospace devices, energy conversion and storage, and flexible electronics.

High-entropy materials (HEMs) are emergent single-phase crystalline alloys or solid solutions typically composed of four or more principal elements. They occupy the vast uncharted central regions of multi-element phase diagrams, which have emerged as an exciting playground for functional materials design and development. Since the first HEM was proposed in 2004, [1,2] several HEM groups have been reported, including metal alloys, [3–7] oxides, [8–11] carbides, [12] borides, [13] nitrides, [14] and sulfides. [15] Some HEMs exhibit extraordinary physical properties beyond that predicted by the expected rule of mixtures, e.g., the superior fracture toughness in the Cantor alloy (CrMnFeCoNi), [16] the large band gap reduction (1.4 eV-2.4 eV) in fluorite-type rare earth high-entropy oxides, [17] the colossal dielectric constant in (MgCoCuNiZn)O, [9] and the low thermal conductivity (approaching the amorphous limit) in (MgCuCoNiSbZn)O. [18] Such remarkable properties have inspired growing interests to explore and design new HEMs for a wide variety of technological applications, not only structural (e.g., aerospace and transportation), but also energy storage and conversion (e.g., batteries, capacitors, and catalysts), [19,20] and thermal and environmental protections against wear and oxidation/corrosion. [21,22]

The design of HEMs is at a very early stage. What element combinations can form a stable single-phase crystalline HEM remains an unresolved challenging question. Thermodynamically, the relative phase stability of a conventional alloy at temperature $T$ is dictated by the Gibbs free energy: $\Delta G_{mix} = \Delta H_{mix} - T\Delta S_{mix}$, where $\Delta H_{mix}$ and $\Delta S_{mix}$ are the enthalpy of mixing and the entropy of mixing, respectively. For chemically ordered crystalline materials, $\Delta S_{mix}$ is often negligible and $\Delta H_{mix}$ dominates the phase stability. The last few decades have witnessed great success in the prediction and design of ordered materials based on $\Delta H_{mix}$ from high-throughput first-principles calculations. [23] However, for HEMs, their design and prediction from first-principles remain formidable. This situation can be attributed to multiple factors. First, the phase stability and formation of HEMs are controlled by the competition between $\Delta H$ and $T\Delta S$. Predicting new HEMs must thus consider $\Delta S$, which is a very difficult quantity to parameterize and estimate from first-principles even with the help of advanced sampling algorithms. [24–26] Second, the number of competing phases of a material increases factorially with the number of its constituent elements. Each HEM often has tens or hundreds of competing phases. Finding those competing phases and assessing their relative stability with respect to the HEM are computationally very demanding even without considering $\Delta S$. Third, there is no unique way to define and model random disorder phases. The supercells required to capture disorder effects are often too large for any practical high-throughput first-principles screening of functional properties over the immense space of candidates.

Existing approaches for exploring new HEMs are mostly based on empirical or semiempirical descriptors or criteria. Commonly used are the Hume-Rothery rules that are based on atomic factors such as atomic sizes, electronegativities, and electron-to-atom ratios. [27,28] These rules are knowingly insufficient as many element combinations that satisfy these rules do not form single-phase solid solutions. [29] Beyond Hume-Rothery rules, two different types of descriptors were also developed. The first type is based on mixing enthalpies combined with/without various atomic factors, assuming an ideal mixing entropy. [30–36] The second type is based on entropy, assuming that $\Delta H$ is universally unimportant. [37] Neither type of these descriptors captures the actuating enthalpy-entropy competition in real materials. As implied from the assumption, the $\Delta H$-based descriptors generally could not clearly separate solid solution phases from intermetallic phases, [32] because the difference in $\Delta H$ can be easily smeared out by the difference in non-ideal $\Delta S$ contributions in real materials. Similarly, the $\Delta S$-based descriptors could not separate crystalline solid-solution phases from amorphous solid-solution phases because both crystalline and amorphous solid solutions can have the same or comparable configurational entropy. In addition to these descriptors, phase diagram methods [38–40] and machine learning [41–45] have also been proposed for predicating some new HEMs, but they depend on the availability of sufficient and reliable experimental data.

Here, we propose a Mixed Enthalpy-Entropy Descriptor (MEED) that features an actuating enthalpy-entropy competition that underlies the stability and formation of real solid solutions. The MEED can be easily and quickly calculated from first-principles, requiring no experimental or empirically derived inputs, and thus enable robust and efficient high-throughput prediction and screening of new HEMs over large chemical spaces. Benchmarking over the carbides and 2D sulfides of metals (Hf, Nb, Mo, Ta, Ti, V, W, and Zr), the MEED successfully identifies all experimentally reported single-phase high-entropy materials in these two groups. By this MEED, these known high-entropy materials are clearly separated from those element combinations that are experimentally known to form multiple phases. A universal cutoff criterion is also identified. With this descriptor, additional new high-entropy carbides and 2D high-entropy transition metal sulfides are also predicted.

***MEED Rationales:*** Our proposed MEED descriptor is based on two distinctive parameters. The first one, denoted as $\Delta H_r$, is the enthalpy difference between the solid solution and the most stable compound formed by its constituent elements (Fig. 1a). The second parameter, denoted as $\langle|\Delta E_D|\rangle$, is the average of absolute differences in defect formation energy among all possible substitutional point defects in the respective constituent binaries that have the same structure as the HEM. These two parameters, taken together, account for the material-specific actuating enthalpy-entropy competition that determines the stability and formation of solid solution phases. They are rationalized as follows:

(#1) The synthesizability of a solid solution is critically dependent on $\Delta H_r$. As the most stable competing compound often is a chemically ordered compound that lacks configurational entropy, the difference in free energy (per atom) between the solid solution and the most stable competing compound is dominated by $\Delta H_r - T\Delta S$, where $\Delta S$ is the configurational entropy of the solid solution. When $\Delta H_r < T\Delta S$, the solid solution is thermodynamically more stable and hence more synthesizable than all its competing compounds individually. When $\Delta H_r > T\Delta S$, the solid solution is thermodynamically less stable than some of its competing compounds. It means that when the solid solution and its most stable competing compound are heated simultaneously, the solid solution phase will decompose first as temperature increases; or when grown under same conditions, those more stable competing phases will precipitate faster than the solid solution. In this sense, the value of $\Delta H_r$ can be viewed as the range of experimental conditions (e.g., temperature and pressure) under which the solid solution is less synthesizable than its most stable competing compound.

(#2) $\Delta H_r$ is better than $\Delta H_{mix}$ in measuring the relative synthesizability among different compounds. It is important to note that $\Delta H_{mix}$ measures the relative enthalpy of the compound with respect to the linear combination of its multiple competing compounds, not to each of these competing compounds individually. A negative $\Delta H_{mix}$ does not mean that this compound is thermodynamically more stable or synthesizable than each of its competing compounds individually. For example, $CsPbI_3$ ($H_f$: −1.27 eV/atom) has a $\Delta H_{mix}$ of −0.02 eV/atom with respect to the combination of competing compounds CsI ($H_f$: −1.74 eV/atom) and $PbI_2$ ($H_f$: −0.92 eV/atom). However, individually, $CsPbI_3$ is less stable or synthesizable than CsI because CsI has a more negative $H_f$. Figure 1b shows our calculated $\Delta H_r$ and $\Delta H_{mix}$ for 28 (8!/6!/2!) ternary solid solutions of carbides $(XY)C_2$ (X, Y= Hf, Nb, Mo, Ta, Ti, V, W, or Zr) with respect to competing phases XC and YC in the same rock-salt structure (cf. Supplemental Fig. S1a). Remarkably, we find that $\Delta H_r$ increases with the increasing $\left|H_f^{\mathrm{XC}} - H_f^{\mathrm{YC}}\right|$. This correlation aligns well with the common wisdom that the larger the difference in $H_f$ (or bond strengths) between two compounds, the more difficult it is to make solid solutions out of them. Such a correlation is not seen in $\Delta H_{mix}$. As shown in Fig. 1b, the $\Delta H_{mix}$ values are actually almost randomly distributed within the narrow range (< 0.15 eV/atom) and hence cannot be used a good descriptor for describing the relative formability of different solid solutions.

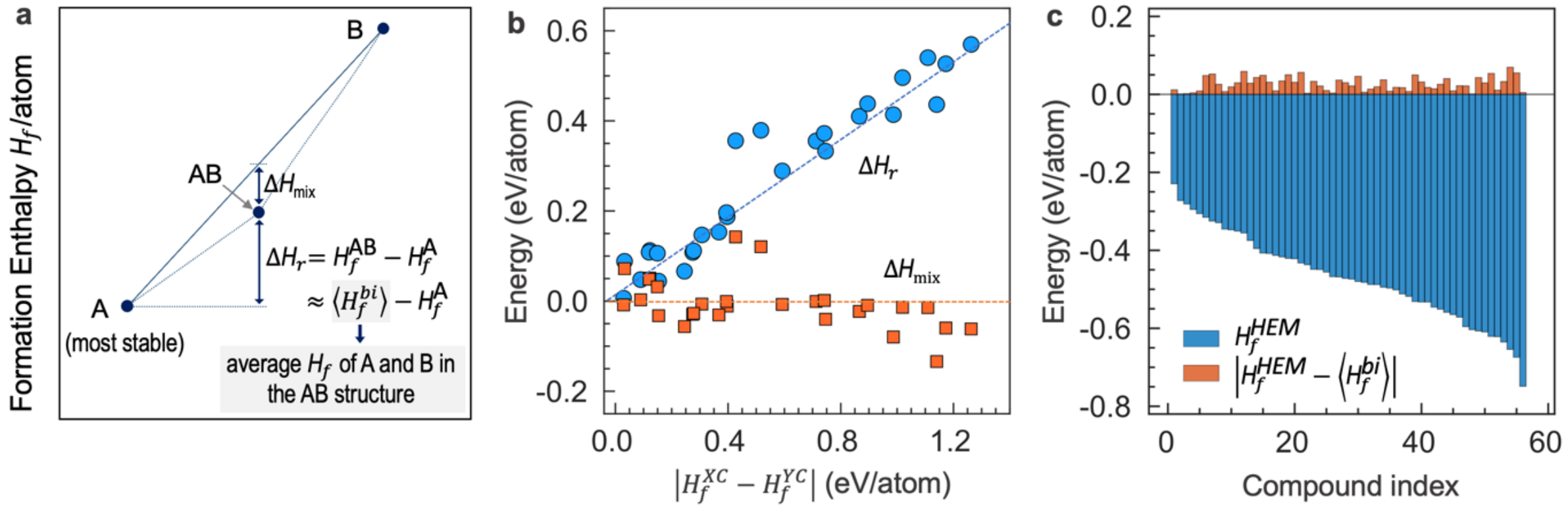

**Figure 1.** (a) The schematic diagram of enthalpy components of an alloy AB and its competing phases A and B. (b) The variation of $\Delta H_r$ and $\Delta H_{\mathrm{mix}}$ of 28 ternary solid solutions of carbides $(XY)C_2$ (X, Y= Hf, Nb, Mo, Ta, Ti, V, W, or Zr) with respect to competing phases XC and YC in the same rock-salt structure. (c) The energy distributions of the heat of formation of 56 five-metal carbides ($H_f^{\mathrm{HEM}}$) and its difference from the average formation enthalpy of the competing binary compounds ($\langle H_f^{\mathrm{bi}} \rangle$) in the same solid solution structure.

(#3) The formation enthalpy of a high-entropy material ($H_f^{\mathrm{HEM}}$) can be approximated with good accuracy by the average formation enthalpy, $\langle H_f^{\mathrm{bi}} \rangle$, of the competing binary compounds in the same solid solution structure. To verify this, we calculated $\left|H_f^{\mathrm{HEM}} - \langle H_f^{\mathrm{bi}} \rangle\right|$ for 56 five-metal carbides in the Special Quasi-random Structure (SQS) [46] (cf. Supplemental Fig S1b). It is found that the $\left|H_f^{\mathrm{HEM}} - \langle H_f^{\mathrm{bi}} \rangle\right|$ is only 26 meV/atom (~ 5% of $H_f^{\mathrm{HEM}}$) in average among in all 56 HEMs (Fig. 1c). With respect to $\langle H_f^{\mathrm{tr}} \rangle$ (the average $H_f$ of the ternary compounds in the same solid solution structure), $\left|H_f^{\mathrm{HEM}} - \langle H_f^{\mathrm{tr}} \rangle\right|$ becomes even smaller (~10 meV/atom) (see Supplemental Fig. S2). Such small differences are consistent with the recent database survey indicating that the recursive enthalpy gain of an *N*-species ordered compound (with respect to the combinations of (*N*-1)-species ordered sub-components) decreases rapidly as *N* increases. [47,48] Recognizing that $H_f^{\mathrm{HEM}} \approx \langle H_f^{\mathrm{bi}} \rangle$ is significant, as it means that $\Delta H_r \approx \langle H_f^{\mathrm{bi}} \rangle - H_f^{\mathrm{min}}$ can be calculated in a highly efficient high-throughput way from the constituent binaries in the same HEM structure and the most stable ordered competing compound (cf. Fig. 1a) without involving large supercells and numerous competing phases.

(#4) In general, the chemical mixing in real materials is not ideal and the configurational entropy of mixing $\Delta S \neq k_B ln(N)$. The $\Delta S$ may vary significantly from one material to another and strongly depend on the chemical composition and crystal structure of the material. In a HEM, principal elements are randomly distributed at different lattice sites of a certain crystal structure. This randomness means that those principal elements can be exchanged or substituted by each other at these lattice sites without causing much energy difference; otherwise, those defects with much lower formation energies will tend to cluster together and form more stable intermetallic phases. From the defect-formation point of view, it means that all possible substitutional point defects in their constituent ordered component materials (usually binaries) in the same crystal structure of the HEM would have small differences in defect formation energy ($E_D$). The degree of randomness or relative magnitude of entropy thus manifests itself in the $E_D$-distribution spectrum of those point defects: the narrower the spectrum, the more random the chemical disorder, and hence the higher the configurational entropy. This rational brings up the second parameter of our descriptor: $\langle|\Delta E_D|\rangle$, which measures the broadness of the $E_D$-spectrum by the average of absolute differences in $E_D$ among all possible substitutional point defects in those constituent binaries of the respective HEM.

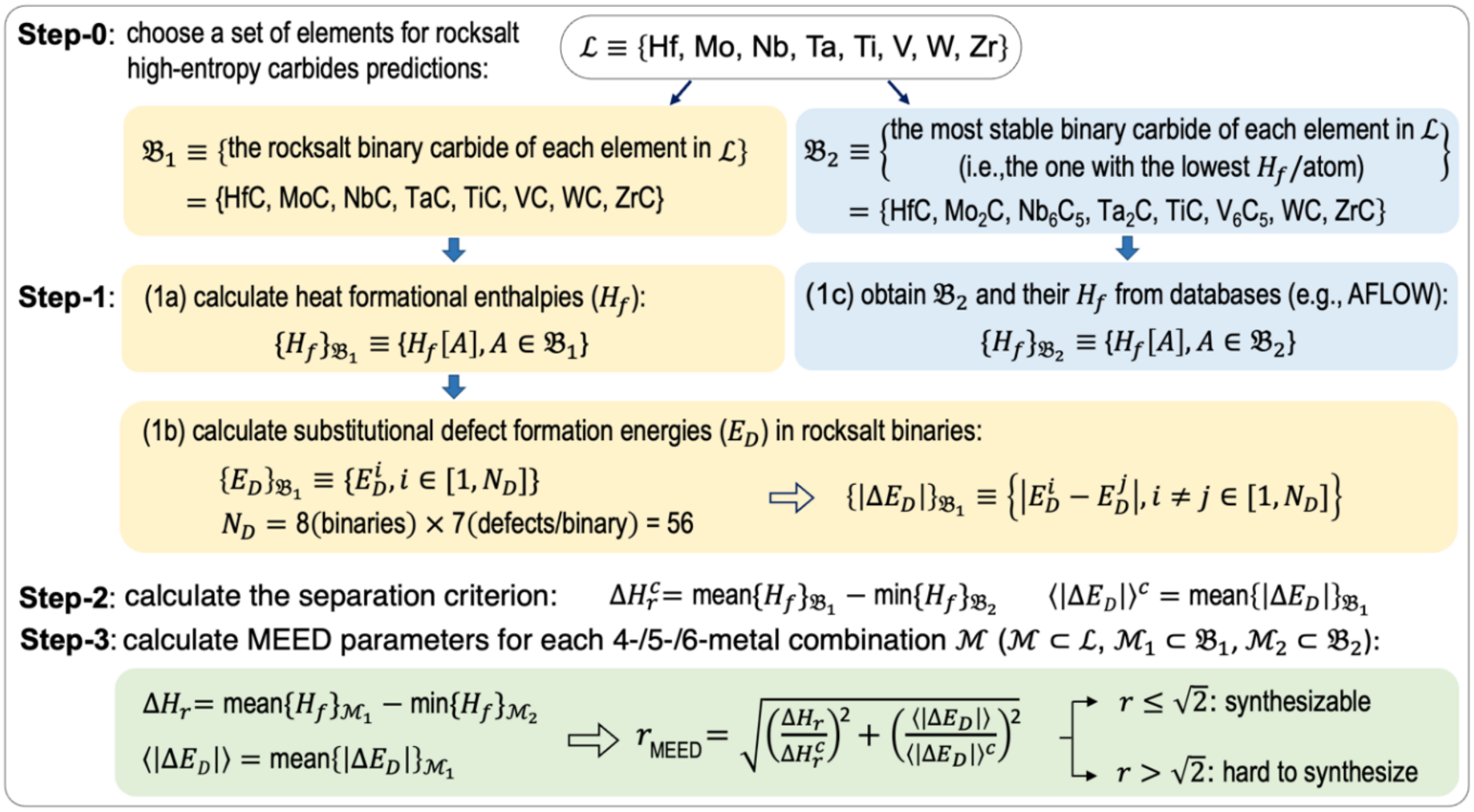

**Figure 2.** The flowchart of calculating MEED parameters for high-entropy metal carbides.

***MEED Benchmark and Predictions:*** To benchmark the MEED, we focus on the rock-salt metal carbide systems that have been extensively studied in experiments. Metal elements are chosen from eight refractory metals (Hf, Nb, Mo, Ta, Ti, V, W, and Zr). Figure 2 shows the flow chart on how MEED parameters are calculated for each element combination from first principles. Details of calculations are given in the Methods section (cf. Supplemental Note S1).

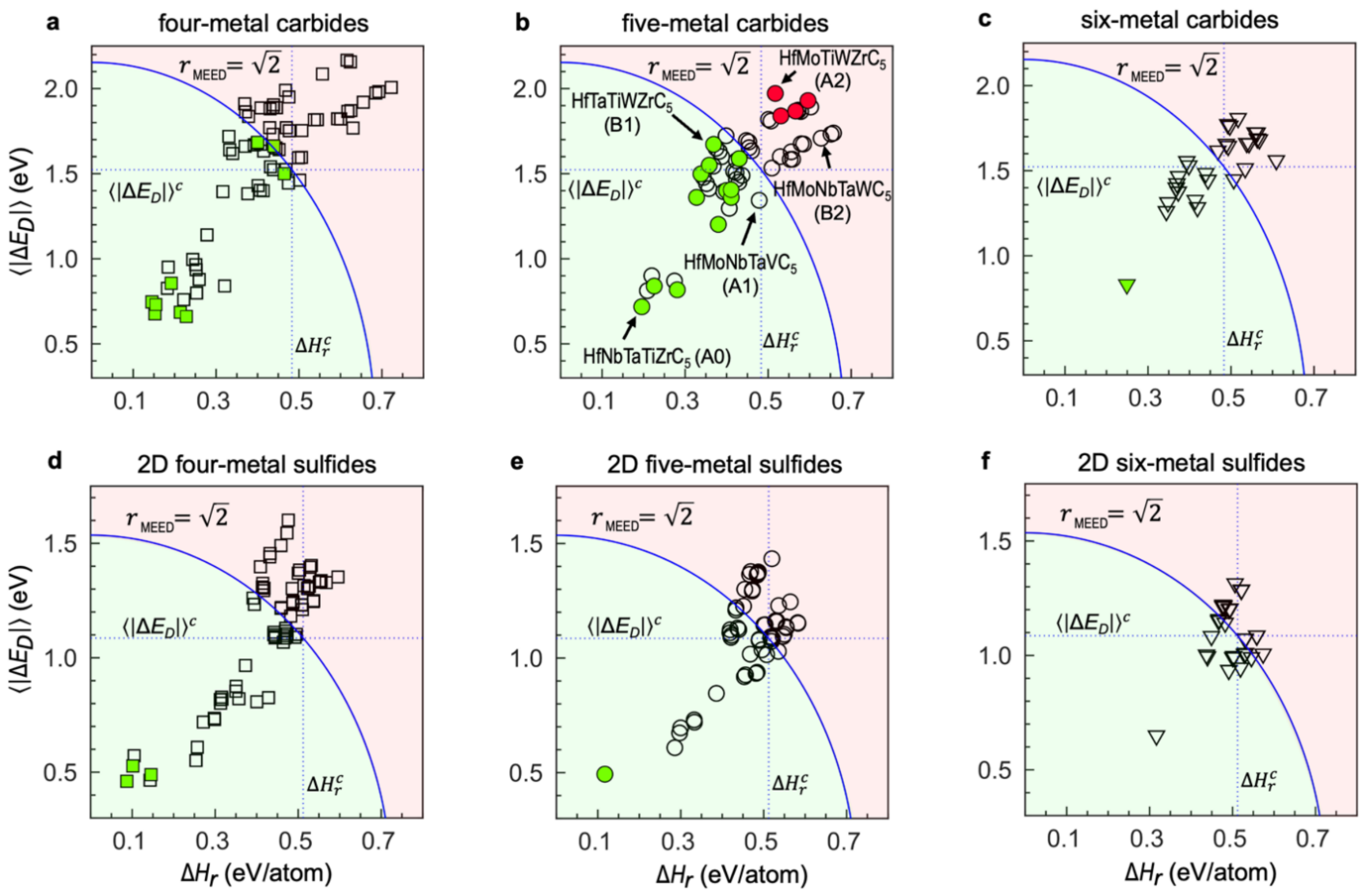

**Figure 3.** Calculated MEED parameters for four-, five-, and six-metal carbides (a-c) and 2D sulfides (d-e) from the eight-metal element set (Hf, Nb, Mo, Ta, Ti, V, W, or Zr). Green symbols represent experimentally synthesized single-phase high-entropy materials. Red symbols represent those experimentally reported multiple-phase compounds.

Fig. 3a-c show our calculated $\Delta H_r$ and $\langle|\Delta E_D|\rangle$ for all 70 (8!/4!/4!) four-metal carbides, 56 (8!/5!/3!) five-metal carbides, and 28 (8!/6!/2!) six-metal carbides that can be generated from the eight-element set. Their associated formation enthalpies of binaries and defect formation energies are listed in Supplemental Table S1. Remarkably, the MEED predicts all presently known single-phase multi-principal metal carbides (marked as green circles) that have been synthesized to date. [12,37,49–60] The closer the solid solution sits to the bottom-left corner of the $\Delta H_r$-$\langle|\Delta E_D|\rangle$ plot, the higher the synthesizability. The predicted top four-metal carbide $HfNbTiZrC_4$, [49] top five-metal carbide $HfNbTaTiZrC_5$, [61,62] and top six-metal carbide $HfNbTaTiVZrC_6$ [60] have all been experimentally validated. As illustrated in Fig. 4a-4b, the top five-metal carbide ($HfNbTaTiZrC_5$) (marked as A0 Fig. 3b) has a much broader $\Delta E_D$-spectrum than the $HfMoTiWZrC_5$ compound (marked as A2 in Fig. 3b). Interestingly, those experimentally reported four- and five-metal carbides are further clustered into two groups. The first group compounds do not contain Mo and W in their element combinations, and they are lower in both $\Delta H_r$ and $\langle|\Delta E_D|\rangle$ (and hence more synthesizable) than the second group compounds which contain elements Mo and/or W. Notably, Mo- and W- carbides are the only two binary compounds that do not have the same cubic rock-salt structure as other binaries. Therefore, as expected, the first group compounds are more synthesizable because their respective binaries exist in the same solid solution structure (consistent with Hume-Rothery rules), and due to this reason, it is easy to amalgamate them during the experimental growth. This structural-chemical aspect is intrinsically incorporated in the MEED.

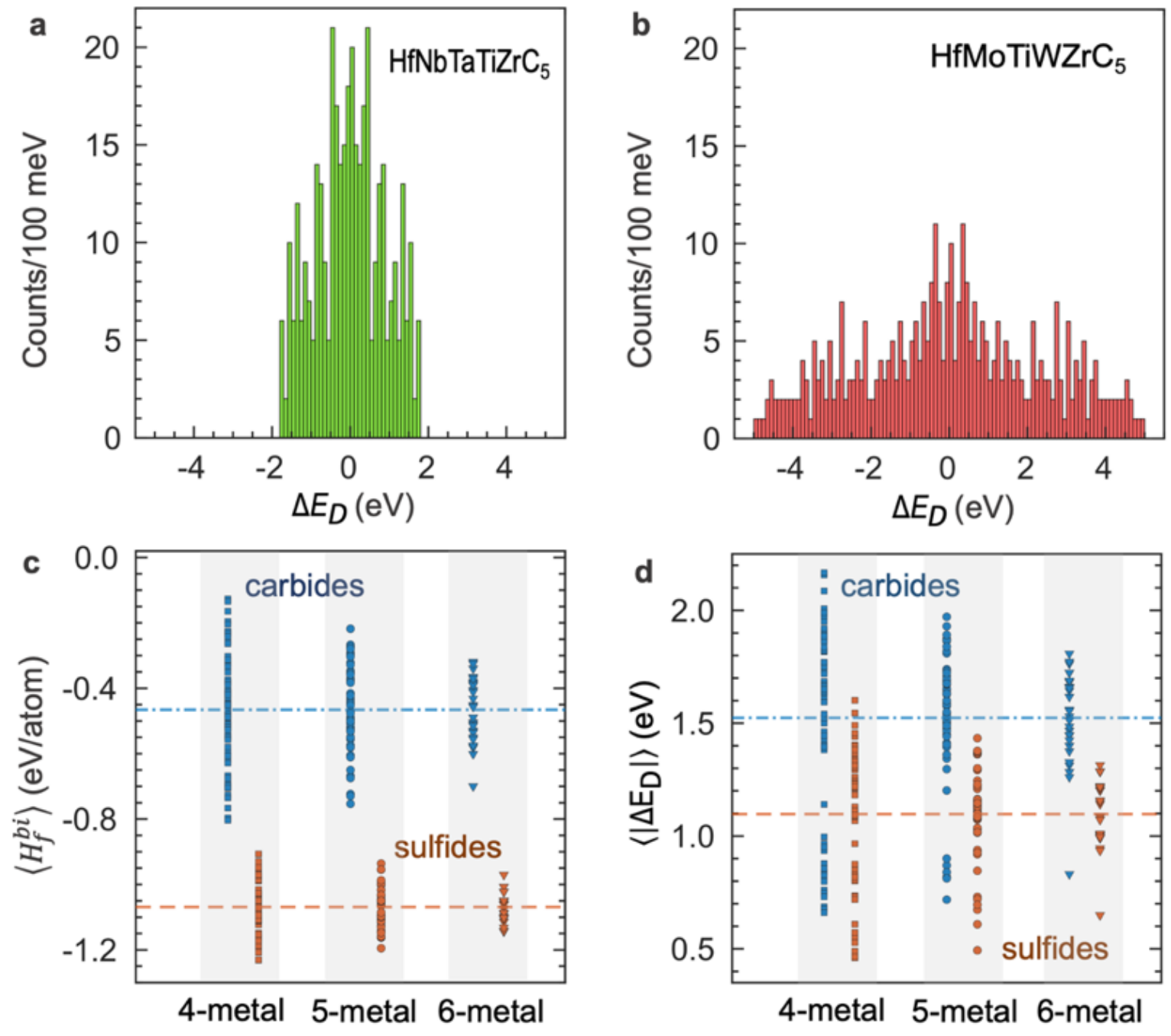

**Figure 4.** Calculated $\Delta E_D$-distribution spectra for experimentally reported single phase compound HfNbTaTiZrC5 (a) and multiphase compound HfMoTiWZrC5 (b). The distribution of (c) $\langle H_f^{\mathrm{bi}} \rangle$ and (d) $\langle |\Delta E_D| \rangle$ for all four-, five-, and six-metal carbides (blue symbols) and sulfides (orange symbols). The average values of the respective parameters for each set of compounds are shown as dotted lines.

More remarkably, as shown in Fig. 3b, all 12 experimentally known single-phase five-metal carbides are clearly separated in both $\Delta H_r$ and $\langle |\Delta E_D| \rangle$ from those experimentally known multiphase ones (marked by red circles). [37,56–59,63] The separation line in the figure is given by:

$$r_{\mathrm{MEED}}(\Delta H_r, \langle |\Delta E_D| \rangle) \equiv \sqrt{\left(\frac{\Delta H_r}{\Delta H_r^c}\right)^2 + \left(\frac{\langle |\Delta E_D| \rangle}{\langle |\Delta E_D| \rangle^c}\right)^2} = \sqrt{2},$$

where $\Delta H_r^c$ is the average $H_f$ of all eight binaries in the rock-salt structure relative to the $H_f^{\mathrm{min}}$ of all eight binary carbides, and $\langle |\Delta E_D| \rangle^c$ is the average of absolute differences in $E_D$ among all 56 substitutional defects. It is worth highlighting that $\Delta H_r^c$ and $\langle |\Delta E_D| \rangle^c$ are two global quantities defined over the whole set of elements under study and are independent of sub-element combinations. The $r_{\mathrm{MEED}}$ is a dimensionless descriptor that unifies $\Delta H_r$ and $\langle |\Delta E_D| \rangle$. Clearly, for a given chemical system, the $r_{\mathrm{MEED}}$ cutoff is set by the values of $\Delta H_r^c$ and $\langle |\Delta E_D| \rangle^c$ and the $r_{\mathrm{MEED}} = \sqrt{2}$ cutoff line passes through the point $(\Delta H_r^c, \langle |\Delta E_D| \rangle^c)$. The smaller the $r_{\mathrm{MEED}}$, the more synthesizable the solid solution. The solid solutions with $r_{\mathrm{MEED}} \leq \sqrt{2}$ are considered synthesizable whereas those with $r_{\mathrm{MEED}} > \sqrt{2}$ are considered hardly synthesizable if not un-synthesizable.

**Table 1.** Calculated values of enthalpy-component ($\Delta H_r$), entropy-component ($\langle|\Delta E_D|\rangle$) of MEED, and $r_{\mathrm{MEED}}$ for top 10 four-, five-, and six-metal carbide and sulfide systems arranged in ascending order of $r_{\mathrm{MEED}}$. The experimentally synthesized single-phase solid solutions are marked as (s).

| **4-metal carbide** | **$\Delta H_r$ (eV/atom)** | **$\langle\lvert\Delta E_D\rvert\rangle$ (eV)** | **$r_{\mathrm{MEED}}$** | **5-metal carbide** | **$\Delta H_r$ (eV/atom)** | **$\langle\lvert\Delta E_D\rvert\rangle$ (eV)** | **$r_{\mathrm{MEED}}$** | **6-metal carbide** | **$\Delta H_r$ (eV/atom)** | **$\langle\lvert\Delta E_D\rvert\rangle$ (eV)** | **$r_{\mathrm{MEED}}$** |
|---|---|---|---|---|---|---|---|---|---|---|---|
| HfNbTiZrC$_4$ (s) [49] | 0.15 | 0.68 | 0.55 | HfNbTaTiZrC$_5$ (s) [37,56,61,62] | 0.20 | 0.72 | 0.62 | HfNbTaTiVZrC$_6$ (s) [60] | 0.25 | 0.83 | 0.75 |
| HfTaTiZrC$_4$ (s) [12] | 0.15 | 0.75 | 0.57 | NbTaTiVZrC$_5$ | 0.21 | 0.81 | 0.69 | MoNbTaTiVZrC$_6$ | 0.34 | 1.26 | 1.09 |
| NbTaTiZrC$_4$ (s) [53] | 0.15 | 0.73 | 0.58 | HfNbTiVZrC$_5$ (s) [58] | 0.23 | 0.84 | 0.72 | HfMoNbTaTiZrC$_6$ | 0.35 | 1.31 | 1.12 |
| HfNbTaZrC$_4$ (s) [12] | 0.21 | 0.69 | 0.63 | HfTaTiVZrC$_5$ | 0.22 | 0.90 | 0.75 | HfMoNbTiVZrC$_6$ | 0.37 | 1.38 | 1.19 |
| NbTaTiVC$_4$ (s) [52] | 0.23 | 0.66 | 0.64 | HfNbTaTiVC$_5$ (s) [37] | 0.28 | 0.82 | 0.79 | HfMoTaTiVZrC$_6$ | 0.37 | 1.40 | 1.19 |
| HfTiVZrC$_4$ | 0.18 | 0.83 | 0.66 | HfNbTaVZrC$_5$ | 0.27 | 0.87 | 0.81 | NbTaTiVWZrC$_6$ | 0.37 | 1.42 | 1.21 |
| HfNbTaTiC$_4$ | 0.22 | 0.76 | 0.68 | MoNbTaTiVC$_5$ (s) [56] | 0.38 | 1.20 | 1.11 | HfMoNbTaTiVC$_6$ | 0.42 | 1.28 | 1.21 |
| NbTiVZrC$_4$ (s) [55] | 0.19 | 0.86 | 0.69 | MoNbTaTiZrC$_5$ (s) [59] | 0.33 | 1.36 | 1.12 | HfMoNbTaVZrC$_6$ | 0.41 | 1.33 | 1.22 |
| TaTiVZrC$_4$ | 0.18 | 0.95 | 0.73 | MoNbTiVZrC$_5$ | 0.36 | 1.41 | 1.19 | HfNbTaTiWZrC$_6$ | 0.37 | 1.47 | 1.24 |
| NbTaVZrC$_4$ | 0.25 | 0.80 | 0.74 | MoTaTiVZrC$_5$ | 0.35 | 1.44 | 1.19 | HfNbTiVWZrC$_6$ | 0.40 | 1.53 | 1.30 |
| **4-metal sulfide** | **$\Delta H_r$ (eV/atom)** | **$\langle\lvert\Delta E_D\rvert\rangle$ (eV)** | **$r_{\mathrm{MEED}}$** | **5-metal sulfide** | **$\Delta H_r$ (eV/atom)** | **$\langle\lvert\Delta E_D\rvert\rangle$ (eV)** | **$r_{\mathrm{MEED}}$** | **6-metal sulfide** | **$\Delta H_r$ (eV/atom)** | **$\langle\lvert\Delta E_D\rvert\rangle$ (eV)** | **$r_{\mathrm{MEED}}$** |
| MoNbTaVS$_8$ (s) [64] | 0.09 | 0.46 | 0.46 | MoNbTaVWS$_{10}$ (s) [64] | 0.12 | 0.49 | 0.51 | MoNbTaTiVWS$_{12}$ | 0.32 | 0.65 | 0.86 |
| MoTaVWS$_8$ | 0.14 | 0.46 | 0.51 | MoNbTaTiVS$_{10}$ | 0.29 | 0.61 | 0.79 | HfNbTaTiVZrS$_{12}$ | 0.44 | 0.99 | 1.26 |
| MoNbTaWS$_8$ (s) [64] | 0.10 | 0.53 | 0.52 | MoNbTaTiWS$_{10}$ | 0.30 | 0.67 | 0.85 | HfMoNbTaTiZrS$_{12}$ | 0.44 | 1.00 | 1.26 |
| MoNbVWS$_8$ (s) [64] | 0.15 | 0.49 | 0.53 | NbTaTiVWS$_{10}$ | 0.30 | 0.70 | 0.87 | MoNbTaTiVZrS$_{12}$ | 0.49 | 0.94 | 1.29 |
| NbTaVWS$_8$ | 0.10 | 0.57 | 0.57 | MoTaTiVWS$_{10}$ | 0.33 | 0.72 | 0.93 | HfNbTaTiWZrS$_{12}$ | 0.45 | 1.08 | 1.33 |
| MoNbTaTiS$_8$ | 0.25 | 0.55 | 0.71 | MoNbTiVWS$_{10}$ | 0.33 | 0.73 | 0.93 | HfMoNbTaTiVS$_{12}$ | 0.52 | 0.95 | 1.34 |
| NbTaTiVS$_8$ | 0.26 | 0.61 | 0.75 | HfNbTaTiZrS$_{10}$ | 0.39 | 0.85 | 1.08 | MoNbTaTiWZrS$_{12}$ | 0.50 | 0.99 | 1.34 |
| NbTaTiWS$_8$ | 0.27 | 0.72 | 0.85 | MoNbTaTiZrS$_{10}$ | 0.45 | 0.92 | 1.22 | NbTaTiVWZrS$_{12}$ | 0.50 | 1.00 | 1.34 |
| MoTaTiVS$_8$ | 0.30 | 0.73 | 0.89 | NbTaTiVZrS$_{10}$ | 0.46 | 0.93 | 1.23 | HfMoNbTaTiWS$_{12}$ | 0.53 | 1.01 | 1.39 |
| MoNbTiVS$_8$ | 0.30 | 0.74 | 0.89 | HfMoNbTaTiS$_{10}$ | 0.48 | 0.93 | 1.27 | HfNbTaTiVWS$_{12}$ | 0.53 | 1.01 | 1.39 |

Our calculated $\Delta H_r$, $\langle|\Delta E_D|\rangle$, and $r_{\mathrm{MEED}}$ for all considered metal carbides are summarized in Supplemental Tables S2-S4. There are 37 four-metal carbides, 32 five-metal carbides, and 13 six-metal carbides that are predicted with $r_{\mathrm{MEED}} \leq \sqrt{2}$. Sixty of them have not been experimentally reported so far. The top ten candidates in each group are listed in Table 1. Disordered metal carbides are known by their super-hard mechanical properties and high melting temperatures. These new high-entropy carbides predicted by MEED offer an avenue to experimentally explore and design metal carbides with enhanced thermo-mechanical properties that are desired for thermal protection coating in aerospace applications and drill bits and cutting tools in mining and industry.

We now apply the MEED to predict new layered 2D high-entropy transition metal sulfides that has been attracting increasing interest. Here we focus on the 2H transition metal sulfides constructed from the same set of metal elements as the carbides. Within this element set, only one five-meal sulfide ($MoNbTaVWS_{10}$) and three four-metal sulfides ($MoNbTaVS_8$, $MoNbVWS_8$, and $MoNbTaWS_8$) have been experimentally reported so far. [64] Our calculated MEED results for 2D sulfides are shown in Fig. 3d-f and Supplemental Tables S6-S8. Their associated formation enthalpies of binaries and defect formation energies are listed in Supplemental Table S5. Remarkably, all these experimentally reported 2D transition metal sulfides (highlighted in color green) turn out to be the top candidates of the MEED prediction. These predicted 2H crystal structure and compositions also agree well with experiments. With the same cutoff criterion $r_{\mathrm{MEED}} = \sqrt{2}$, the MEED predicts 31 new four-metal sulfides, 25 new five-metal sulfides, and 14 new six-metal sulfides. The top ten predicted high entropy 2D metal sulfides are also listed in Table 1. These new HEMs with combined features of "high-entropy" and "reduced dimensionality (2D) and atomic thinness" may display unusual physical properties and be promising for many technologically important applications such as energy conversion and storage, catalysts, and flexible electronics.

***Emerging Trends and Design Rules:*** A common trend displayed in Fig. 3 is that $\langle|\Delta E_D|\rangle$ generally increases with increasing $\Delta H_r$. The solid solution with a small $\Delta H_r$ is generally also small in $\langle|\Delta E_D|\rangle$, and vice versa. In other words, the closer the enthalpy of formation between the solid solution and its most stable competing compound, the narrower the spread of the defect formation energy spectrum, the more random the chemical disorder, and hence the higher the entropy of the solid solution. This trend brings up a general design rule for the HEM prediction, that is, the element combination with a smaller $\langle|\Delta E_D|\rangle$ or $\Delta H_r$ in a certain crystal structure can more easily form into a single-phase solid solution.

A more important trend embodied in Fig. 3 is that the increasing correlation between $\Delta H_r$ and $\langle|\Delta E_D|\rangle$ is rather scattered. For a same or similar $\Delta H_r$ ($\langle|\Delta E_D|\rangle$), the $\langle|\Delta E_D|\rangle$ ($\Delta H_r$) can still vary significantly from one compound to another. For example, as shown in Fig. 3b, $HfMoNbTaVC_5$ (marked as A1) and $HfMoTiWZrC_5$ (marked as A2) have only 39 meV difference in $\Delta H_r$ but their difference in $\langle|\Delta E_D|\rangle$ is 16 times larger (628 meV); $HfTaTiWZrC_5$ (marked as B1) and $HfMoNbTaWC_5$ (marked as B2) have only 36 meV difference in $\langle|\Delta E_D|\rangle$ but their difference in $\Delta H_r$ is as large as 259 meV. Hence, the $\Delta H_r$ alone is insufficient to determine the relative synthesizability between A1 and A2 and the $\langle|\Delta E_D|\rangle$ alone is insufficient to determine the relative synthesizability between B1 and B2. The dimensionless MEED descriptor $r_{\mathrm{MEED}}$, which simultaneously into account both $\Delta H_r$ and $\langle|\Delta E_D|\rangle$, nicely distinguishes their relative synthesizability. The cutoff $r_{\mathrm{MEED}} = \sqrt{2}$ is universal and can be applied to other materials systems not studied in this work.

In addition to the dependence of synthesizability on chemical compositions, the MEED also distinguishes the roles of crystal structures. In general, solid solutions with the same chemical composition have different stability and synthesizability in different structures, and the solid solutions in different crystal structures have different element combinations that are most stable or synthesizable in each specific structure. This is true according to the MEED. Table 1 indicates that the MEED-predicted top five-metal HEMs from the same eight-metal element set are different in the rock-salt and 2H layer structures (e.g., $HfNbTaTiZrC_5$ vs $MoNbTaVWS_{10}$). Further, the rock-salt carbides are also found systematically higher in $\Delta H_r$ than the 2H-layer sulfides. It suggests that the 2H-layer sulfides can be synthesized at relatively lower temperatures than the rock-salt carbides. The experimentally reported growth temperatures for rock-salt carbides and 2H-layer sulfides are 1600ºC-2200ºC [37,61] and 1000ºC, [64] respectively, agreeing well with the MEED prediction.

Figure 3 also show that the number of MEED-predicted top candidates and the synthesizability of these candidates decrease with increasing number of constituent metal elements (*N*). This trend is understandable and can be seen more clearly in Fig. 4c-d. For both carbides and sulfides, it is found that the averages in $\langle|\Delta E_D|\rangle$ and $\langle H_f^{\text{bi}}\rangle$ of all possible solid solutions with the same *N* do not change with *N*, while the range of their variations in $\langle|\Delta E_D|\rangle$ and $\langle H_f^{\text{bi}}\rangle$ decreases with increasing *N*. It means that the solid solutions with $\langle|\Delta E_D|\rangle$ and $\langle H_f^{\text{bi}}\rangle$ lower than their respective average values shift up as *N* increases. The up shift in $\langle|\Delta E_D|\rangle$ suggests that creating chemical disorder becomes harder and the actual entropy of mixing does not necessarily increase with increasing *N*. The up shift in $\langle H_f^{\text{bi}}\rangle$ implies a larger $\Delta H_r$ (because the most stable competing compound remains same as *N* increases) and thus a lower stability and synthesizability with respect to its most stable competing compound.

***Discussions:*** The most distinctive feature of the MEED lies in how enthalpy and entropy are defined and quantified to assess the stability and synthesizability of HEMs. Conventionally, the enthalpy contribution to the stability of a HEM is evaluated from the convex hull of ~$10^2$-$10^3$ of competing phases of the HEM in the SQS structure. Here we find that (i) the formation enthalpy of a HEM can be approximated in good accuracy by the average formation energies of only 4 to 6 ordered binary compounds and (ii) the synthesizability is correlated to the formation enthalpy of the HEM relative to the "most stable" competing phase rather than to its "all possible" competing phases. For the entropy contribution, the common estimation methods are based on the sampling of alloy configurations, [35–37] assuming that the HEM phase can be viewed as a linear combination of ~$10^2$-$10^4$ of ordered structures or clusters. Here we find that (iii) the entropy forming ability can be determined from only tens of DFT-calculated point defect-formation energies. These findings are remarkable and important as they allow one to assess the stability and synthesizability of HEMs without considering numerous competing phases and without using large pseudo-random supercell structures.

Another important feature is that the MEED can be easily calculated from DFT, and the computational cost is fractional comparing with other DFT-based methods. For example, to predict the high entropy carbides of five and six metals out of an eight-element set, 17,360 DFT total energy calculations are needed by the EFA (entropy forming ability) descriptor, [37] while by the MEED, only 56 DFT-calculated defect formation energies and 16 DFT-calculated formation enthalpies of binaries (Table S1) are needed. This feature renders the descriptor capable of predicting and screening new HEMs over vast chemical spaces in a high-throughput way.

The dimensionless parameter $r_{\text{MEED}}$, defined over $\Delta H_r$ and $\langle|\Delta E_D|\rangle$, embodies a universal criterion ($r_{\text{MEED}} = \sqrt{2}$) for new HEM prediction. The prediction by this $r_{\text{MEED}}$ goes beyond the capability of existing (semi)empirical descriptors. For instance, the experimentally reported single-phase $MoNbTaVWC_5$ compound, which does not follow Hume-Rothery rules, is successfully predicted here by the MEED with $r_{\text{MEED}} = 1.23$. The chemical trends of those experimentally reported single-phase high-entropy carbides, which are unseen by the EFA descriptor, are clearly shown in $r_{\text{MEED}}$ (Fig.3 and Supplemental Fig.S3). Notably, the predicted "top high-entropy carbides of four, five, and six metals" and "top high-entropy 2D sulfides of four and five metals" have all been experimentally synthesized (Fig.3). These predictions are remarkable given that the carbides and sulfides studied here are of very different crystal structures and dimensionalities (3D cubic rocksalt vs layered 2D hexagonal).

Finally, it may be worth noting that like all other descriptors, the MEED does not consider kinetics. It is assumed here that the kinetic does not play a decisive role in the synthesizability at high temperatures. The vibrational contribution to the stability is not considered here either as it is generally one order of magnitude smaller than the configuration entropy. [40,65] At most, the vibrational contributions may be comparable to the contributions of enthalpy of mixing $\Delta H_{mix}$ in some cases where the solid solution and competing phases have very different structures. However, in such cases, the $\Delta H_r$ is generally also large as found in Fig. 1b and thus the synthesizability is still more dictated by $\Delta H_r$.

**Acknowledgements**

This work has been supported by the U.S. DOE, Office of Science, Office of Basic Energy Sciences, under Award No. DE-SC0021127. D.D. and L.Y. gratefully acknowledge the assistance of the Advanced Computing Group of the University of Maine System for providing computational resources for this work. Part of this research was conducted at the Center for Nanophase Materials Sciences using the resources of the Compute and Data Environment for Science (CADES) at the Oak Ridge National Laboratory, which is a DOE Office of Science User Facility.

# Supplemental Materials for
# Descriptor-Enabled Rational Design of High-Entropy Materials Over Vast Chemical Spaces

Dibyendu Dey[1, *], Liangbo Liang[2], and Liping Yu[1, *]
[1]Department of Physics and Astronomy, University of Maine, Orono, Maine 04469, USA.
[2]Center for Nanophase Materials Sciences, Oak Ridge National Laboratory, Oak Ridge, Tennessee 37831, USA.

**Figure S1.** Special quasi-random structures (SQS) used in first-principles calculations for disordered alloys: (a) 2x2x1 supercell for two-metal (ternary) and (b) 2x2x5 supercell for five-metal carbides; (c) 4x4x1 supercell for two-metal and (d) 5x13x1 supercell for five-metal 2D sulfides.

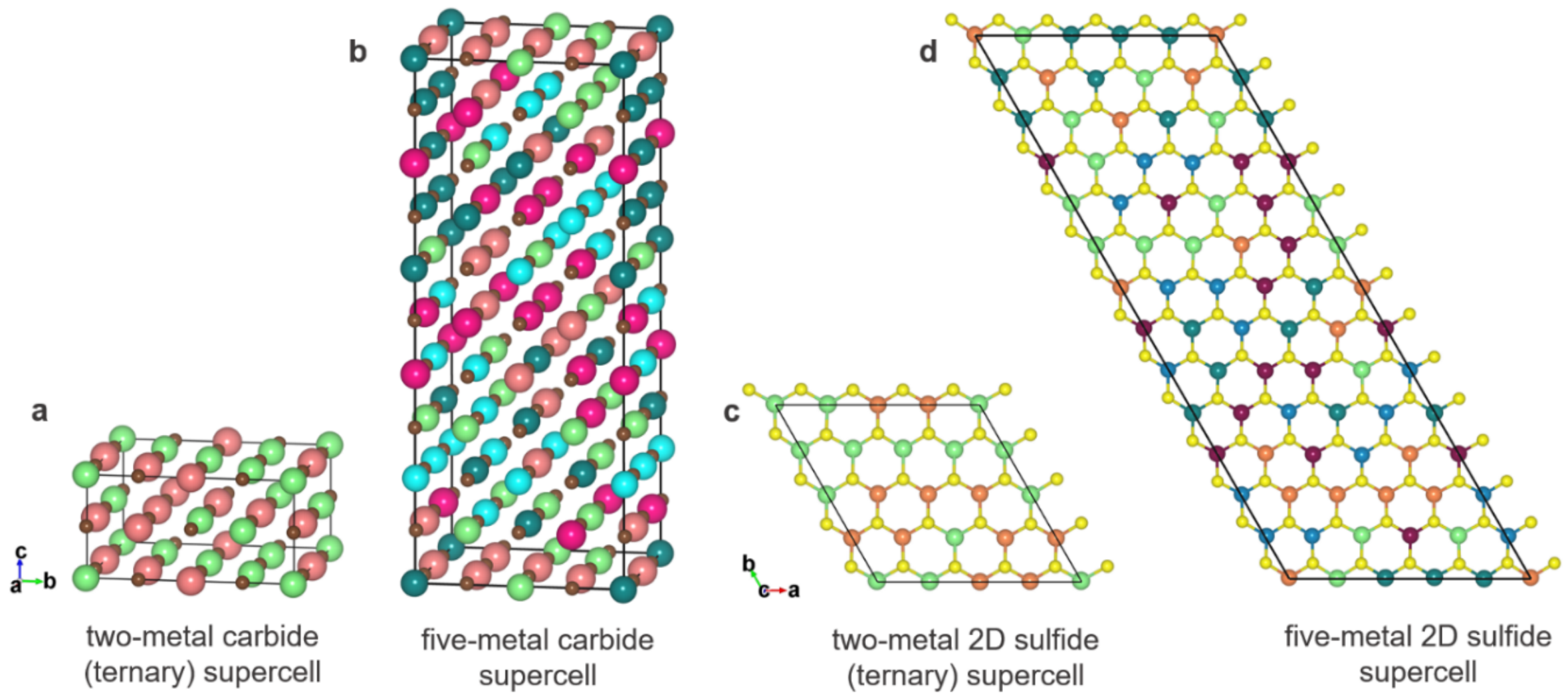

**Figure S2.** The energy distribution of the heat of formation of 56 five-metal carbides ($H_f^{\mathrm{HEM}}$) and its difference from the average formation enthalpy, $\langle H_f^{\mathrm{tr}} \rangle$, of the competing ternary compounds in the same solid solution structure. $\left| H_f^{\mathrm{HEM}} - \langle H_f^{\mathrm{tr}} \rangle \right|$ is only 10 meV/atom in average among in all 56 HEMs.

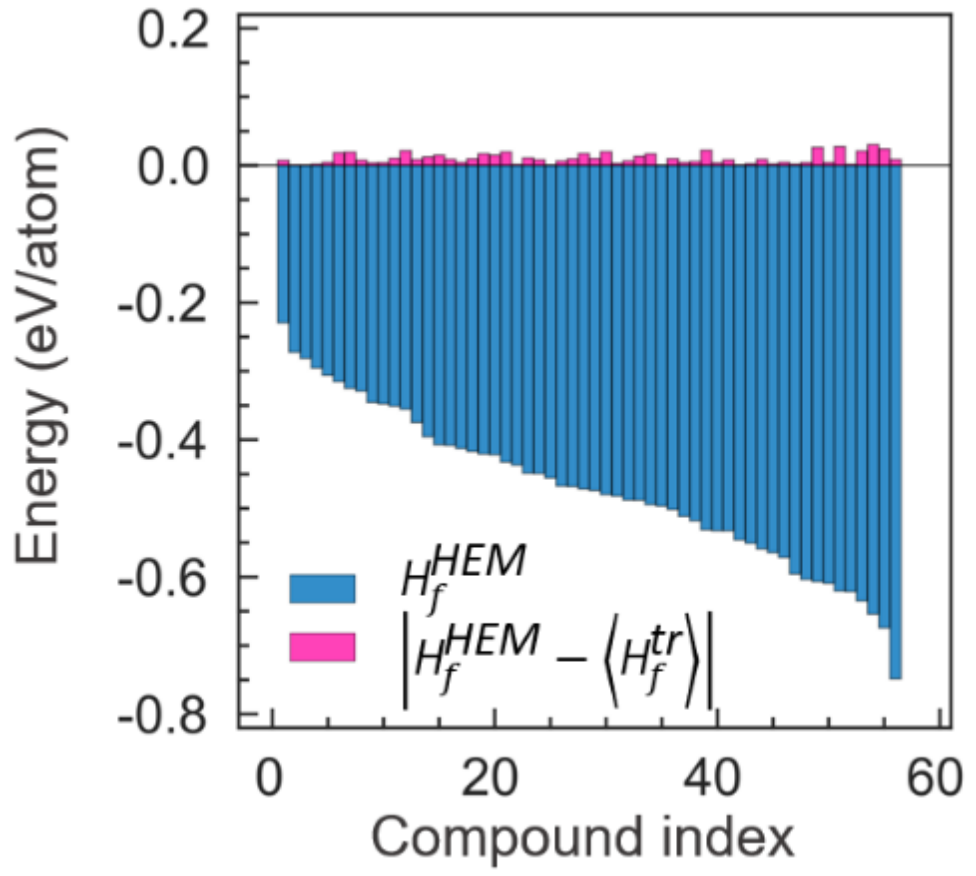

**Figure S3. Comparison between the entropy forming ability (EFA) descriptor and the MEED.** (a) 1/EFA and (b) $r_{MEED}$ values are shown in an ascending order for all 56 five metal carbides. Green and red circles indicate the experimentally synthesized single-phase and multiple-phase compounds, respectively. Key observations include: (i) $r_{MEED}$ is better than EFA descriptor in separating experimentally known single-phase HEMs from multiphase non-high entropy materials; (ii) $r_{MEED}$ embodies a universal selection criterion; and (iii) the experimentally reported single phases (green dots) are clustered in $r_{MEED}$, following the composition-structure-stability relationship. Another distinctive feature is that $r_{MEED}$ can be easily calculated from DFT and the computational cost of it is fractional comparing with EFA.

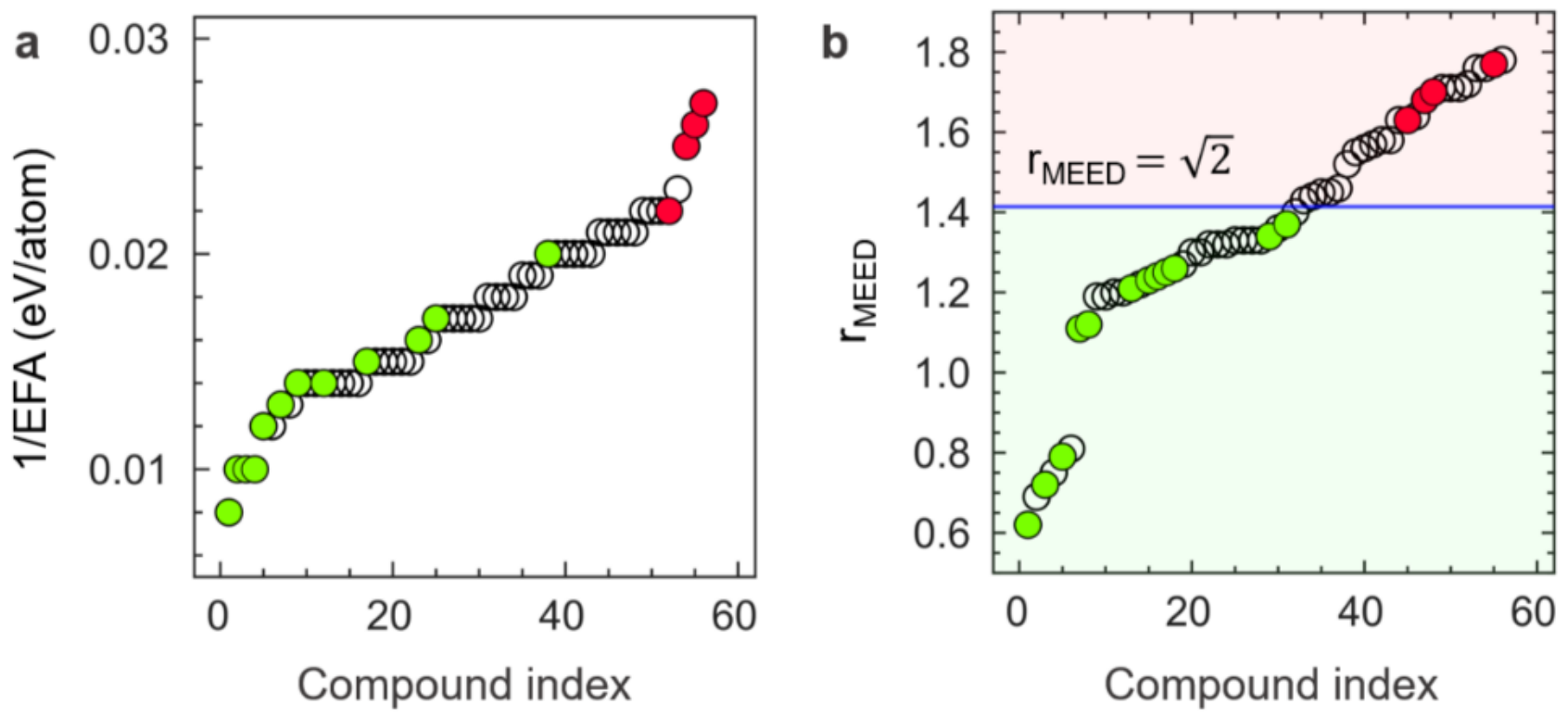

**Table S1.** DFT-calculated formation enthalpies ($H_f$) and defect formation energies ($E_D$) of carbide systems.

| Stablest Binaries | | Rocksalt Binaries | | | | | | | | | |
|---|---|---|---|---|---|---|---|---|---|---|---|
| formula (space group) | (a) $H_f$ (eV/atom) | Formula (#225) | (b) $H_f$ (eV/atom) | (c) $E_D$ of substitutional defects (eV) | | | | | | | |
| | | | | Hf | Mo | Nb | Ta | Ti | V | W | Zr |
| HfC (#225) | -0.95 | HfC | -0.95 | – | 2.02 | 0.49 | 0.25 | 0.60 | 1.88 | 1.84 | 0.17 |
| $Mo_2C$ (#60) | -0.12 | MoC | 0.16 | -2.18 | – | -1.41 | -1.41 | -2.52 | -1.26 | 0.25 | -1.94 |
| $Nb_6C_5$ (#12) | -0.62 | NbC | -0.55 | -0.99 | 1.26 | – | -0.28 | -0.92 | 0.45 | 1.40 | -0.76 |
| $Ta_2C$ (#164) | -0.61 | TaC | -0.58 | -1.18 | 1.21 | -0.29 | – | -1.16 | 0.24 | 1.44 | -0.96 |
| TiC (#225) | -0.83 | TiC | -0.83 | 0.21 | 1.06 | 0.13 | -0.13 | – | 0.66 | 0.84 | 0.60 |
| $V_6C_5$ (#151) | -0.55 | VC | -0.43 | 0.16 | 1.15 | 0.27 | 0.12 | -0.75 | – | 1.19 | 0.60 |
| WC (#187) | -0.13 | WC | 0.31 | -2.83 | -0.36 | -1.88 | -1.81 | -3.19 | -1.83 | – | -2.62 |
| ZrC (#225) | -0.82 | ZrC | -0.86 | -0.18 | 1.93 | 0.38 | 0.15 | 0.58 | 1.90 | 1.73 | – |

**Table S2.** Calculated values of enthalpy-component ($\Delta H_r$), entropy-component ($\langle|\Delta E_D|\rangle$) of MEED, and $r_{MEED}$ for 70 four-metal carbide systems arranged in ascending order of $r_{MEED}$. The experimentally synthesized single-phase solid solutions are marked as (s).

| 4-metal carbide | $\Delta H_r$ (eV/atom) | $\langle\|\Delta E_D\|\rangle$ (eV) | $r_{MEED}$ | 4-metal carbide | $\Delta H_r$ (eV/atom) | $\langle\|\Delta E_D\|\rangle$ (eV) | $r_{MEED}$ | 4-metal carbide | $\Delta H_r$ (eV/atom) | $\langle\|\Delta E_D\|\rangle$ (eV) | $r_{MEED}$ |
|---|---|---|---|---|---|---|---|---|---|---|---|
| HfNbTiZrC$_4$ (s) [1] | 0.15 | 0.68 | 0.55 | MoTiVZrC$_4$ | 0.37 | 1.66 | 1.33 | NbVWZrC$_4$ | 0.48 | 1.75 | 1.52 |
| HfTaTiZrC$_4$ (s) [2] | 0.15 | 0.75 | 0.57 | MoTaVZrC$_4$ | 0.43 | 1.54 | 1.35 | HfTaWZrC$_4$ | 0.43 | 1.88 | 1.52 |
| NbTaTiZrC$_4$ (s) [3] | 0.15 | 0.73 | 0.58 | MoNbVZrC$_4$ | 0.44 | 1.52 | 1.35 | HfNbWZrC$_4$ | 0.44 | 1.89 | 1.54 |
| HfNbTaZrC$_4$ (s) [2] | 0.21 | 0.69 | 0.63 | HfMoTaZrC$_4$ | 0.39 | 1.67 | 1.36 | HfNbTiWC$_4$ | 0.45 | 1.89 | 1.55 |
| NbTaTiVC$_4$ (s) [4] | 0.23 | 0.66 | 0.64 | MoTaVWC$_4$ | 0.47 | 1.44 | 1.37 | HfTaTiWC$_4$ | 0.44 | 1.90 | 1.55 |
| HfTiVZrC$_4$ | 0.18 | 0.83 | 0.66 | NbTaTiWC$_4$ | 0.41 | 1.63 | 1.37 | HfNbTaWC$_4$ | 0.51 | 1.75 | 1.56 |
| HfNbTaTiC$_4$ | 0.22 | 0.76 | 0.68 | MoNbTaWC$_4$ (s) [5] | 0.46 | 1.50 | 1.38 | HfTiVWC$_4$ | 0.48 | 1.95 | 1.62 |
| NbTiVZrC$_4$ (s) [6] | 0.19 | 0.86 | 0.69 | HfMoNbZrC$_4$ | 0.40 | 1.68 | 1.38 | HfVWZrC$_4$ | 0.47 | 1.99 | 1.63 |
| TaTiVZrC$_4$ | 0.18 | 0.95 | 0.73 | HfMoTaTiC$_4$ (s) [7] | 0.40 | 1.68 | 1.38 | HfTaVWC$_4$ | 0.54 | 1.82 | 1.63 |
| NbTaVZrC$_4$ | 0.25 | 0.80 | 0.74 | HfMoNbTiC$_4$ | 0.41 | 1.67 | 1.39 | HfNbVWC$_4$ | 0.54 | 1.82 | 1.64 |
| HfNbTiVC$_4$ | 0.26 | 0.88 | 0.79 | HfMoNbTaC$_4$ | 0.47 | 1.52 | 1.39 | MoTaTiWC$_4$ | 0.59 | 1.82 | 1.71 |
| HfTaTiVC$_4$ | 0.25 | 0.94 | 0.81 | MoNbVWC$_4$ | 0.50 | 1.46 | 1.41 | MoNbTiWC$_4$ | 0.60 | 1.82 | 1.72 |
| HfNbVZrC$_4$ | 0.25 | 0.97 | 0.82 | NbTaWZrC$_4$ (s) [8] | 0.44 | 1.65 | 1.41 | MoTiVWC$_4$ | 0.63 | 1.77 | 1.75 |
| HfTaVZrC$_4$ | 0.24 | 1.00 | 0.83 | TaTiVWC$_4$ | 0.44 | 1.66 | 1.43 | MoTaWZrC$_4$ | 0.62 | 1.87 | 1.77 |
| HfNbTaVC$_4$ | 0.32 | 0.84 | 0.86 | NbTiVWC$_4$ | 0.45 | 1.64 | 1.43 | MoNbWZrC$_4$ | 0.63 | 1.87 | 1.78 |
| MoNbTaVC$_4$ | 0.28 | 1.14 | 0.94 | NbTiWZrC$_4$ | 0.38 | 1.84 | 1.44 | MoTiWZrC$_4$ | 0.56 | 2.09 | 1.79 |
| NbTaVWC$_4$ | 0.32 | 1.39 | 1.13 | TaTiWZrC$_4$ | 0.37 | 1.86 | 1.44 | MoVWZrC$_4$ | 0.66 | 1.92 | 1.85 |
| MoNbTaTiC$_4$ | 0.38 | 1.38 | 1.20 | HfMoTiVC$_4$ | 0.44 | 1.73 | 1.45 | HfMoWZrC$_4$ | 0.62 | 2.17 | 1.91 |
| MoTaTiVC$_4$ | 0.41 | 1.40 | 1.25 | HfMoVZrC$_4$ | 0.43 | 1.77 | 1.46 | HfMoTiWC$_4$ | 0.62 | 2.16 | 1.92 |
| MoNbTaZrC$_4$ | 0.40 | 1.43 | 1.25 | HfTiWZrC$_4$ | 0.37 | 1.91 | 1.47 | HfMoTaWC$_4$ | 0.69 | 1.97 | 1.92 |
| MoNbTiVC$_4$ | 0.41 | 1.40 | 1.26 | HfMoTaVC$_4$ | 0.50 | 1.59 | 1.47 | HfMoNbWC$_4$ | 0.69 | 1.98 | 1.94 |
| MoNbTiZrC$_4$ | 0.34 | 1.62 | 1.27 | HfMoNbVC$_4$ | 0.51 | 1.60 | 1.48 | HfMoVWC$_4$ | 0.72 | 2.01 | 1.99 |
| MoTaTiZrC$_4$ | 0.33 | 1.64 | 1.28 | TiVWZrC$_4$ | 0.41 | 1.89 | 1.50 | | | | |
| HfMoTiZrC$_4$ | 0.33 | 1.72 | 1.32 | TaVWZrC$_4$ | 0.47 | 1.77 | 1.52 | | | | |

**Table S3.** Calculated values of enthalpy-component ($\Delta H_r$), entropy-component ($\langle|\Delta E_D|\rangle$) of MEED, and $r_{MEED}$ for 56 five-metal carbide systems arranged in ascending order of $r_{MEED}$. The experimentally synthesized single-phase solid solutions are marked as (s) and multiple-phase compounds are marked as (m).

| 5-metal carbide | $\Delta H_r$ (eV/atom) | $\langle\|\Delta E_D\|\rangle$ (eV) | $r_{MEED}$ | 5-metal carbide | $\Delta H_r$ (eV/atom) | $\langle\|\Delta E_D\|\rangle$ (eV) | $r_{MEED}$ | 5-metal carbide | $\Delta H_r$ (eV/atom) | $\langle\|\Delta E_D\|\rangle$ (eV) | $r_{MEED}$ |
|---|---|---|---|---|---|---|---|---|---|---|---|
| HfNbTaTiZrC$_5$ (s) [9–12] | 0.20 | 0.72 | 0.62 | HfMoTaTiVC$_5$ | 0.42 | 1.47 | 1.30 | MoTaTiVWC$_5$ | 0.55 | 1.58 | 1.55 |
| NbTaTiVZrC$_5$ | 0.21 | 0.81 | 0.69 | HfMoNbTiVC$_5$ | 0.43 | 1.45 | 1.30 | MoNbTiVWC$_5$ | 0.56 | 1.58 | 1.56 |
| HfNbTiVZrC$_5$ (s) [13] | 0.23 | 0.84 | 0.72 | HfMoTaVZrC$_5$ | 0.42 | 1.51 | 1.32 | MoNbTaWZrC$_5$ | 0.56 | 1.63 | 1.57 |
| HfTaTiVZrC$_5$ | 0.22 | 0.90 | 0.75 | HfMoNbVZrC$_5$ | 0.42 | 1.51 | 1.32 | MoTaTiWZrC$_5$ | 0.50 | 1.82 | 1.58 |
| HfNbTaTiVC$_5$ (s) [10] | 0.28 | 0.82 | 0.79 | NbTiVWZrC$_5$ | 0.39 | 1.60 | 1.32 | MoNbTiWZrC$_5$ | 0.51 | 1.81 | 1.58 |
| HfNbTaVZrC$_5$ | 0.27 | 0.87 | 0.81 | HfMoNbTaVC$_5$ | 0.48 | 1.34 | 1.33 | MoTaVWZrC$_5$ | 0.58 | 1.68 | 1.63 |
| MoNbTaTiVC$_5$ (s) [11] | 0.38 | 1.20 | 1.11 | TaTiVWZrC$_5$ | 0.38 | 1.63 | 1.33 | MoTiVWZrC$_5$ (m) [14] | 0.53 | 1.84 | 1.63 |
| MoNbTaTiZrC$_5$ (s) [15] | 0.33 | 1.36 | 1.12 | HfNbTiWZrC$_5$ | 0.37 | 1.65 | 1.33 | MoNbVWZrC$_5$ | 0.59 | 1.67 | 1.64 |
| MoNbTiVZrC$_5$ | 0.36 | 1.41 | 1.19 | NbTaVWZrC$_5$ | 0.44 | 1.49 | 1.33 | HfMoTiWZrC$_5$ (m) [10] | 0.52 | 1.97 | 1.68 |
| MoTaTiVZrC$_5$ | 0.35 | 1.44 | 1.19 | HfTaTiWZrC$_5$ (s) [10] | 0.37 | 1.67 | 1.34 | HfMoTaWZrC$_5$ (m) [10] | 0.57 | 1.87 | 1.70 |
| MoNbTaVZrC$_5$ | 0.41 | 1.30 | 1.20 | HfNbTaWZrC$_5$ | 0.42 | 1.58 | 1.36 | HfMoTaTiWC$_5$ | 0.57 | 1.87 | 1.71 |
| HfMoNbTiZrC$_5$ | 0.34 | 1.48 | 1.20 | HfNbTaTiWC$_5$ (s) [10] | 0.43 | 1.59 | 1.37 | HfMoNbWZrC$_5$ | 0.57 | 1.87 | 1.71 |
| HfMoTaTiZrC$_5$ (s) [11] | 0.34 | 1.50 | 1.21 | HfTiVWZrC$_5$ | 0.40 | 1.72 | 1.40 | HfMoNbTiWC$_5$ | 0.58 | 1.87 | 1.71 |
| HfMoNbTaZrC$_5$ | 0.39 | 1.39 | 1.22 | HfNbTiVWC$_5$ | 0.46 | 1.63 | 1.43 | HfMoNbTaWC$_5$ | 0.63 | 1.71 | 1.72 |
| MoNbTaVWC$_5$ (s) [10] | 0.41 | 1.36 | 1.23 | HfTaTiVWC$_5$ | 0.45 | 1.65 | 1.44 | HfMoTiVWC$_5$ | 0.60 | 1.89 | 1.76 |
| HfMoNbTaTiC$_5$ (s) [16] | 0.40 | 1.40 | 1.24 | HfTaVWZrC$_5$ | 0.45 | 1.69 | 1.45 | HfMoTaVWC$_5$ | 0.65 | 1.73 | 1.76 |
| NbTaTiVWC$_5$ (s) [10] | 0.41 | 1.41 | 1.25 | HfNbVWZrC$_5$ | 0.45 | 1.69 | 1.45 | HfMoVWZrC$_5$ (m) [10] | 0.60 | 1.93 | 1.77 |
| NbTaTiWZrC$_5$ (s) [17] | 0.36 | 1.55 | 1.26 | HfNbTaVWC$_5$ | 0.51 | 1.53 | 1.46 | HfMoNbVWC$_5$ | 0.66 | 1.74 | 1.78 |
| HfMoTiVZrC$_5$ | 0.37 | 1.54 | 1.27 | MoNbTaTiWC$_5$ | 0.53 | 1.60 | 1.52 | | | | |

**Table S4.** Calculated values of enthalpy-component ($\Delta H_r$), entropy-component ($\langle|\Delta E_D|\rangle$) of MEED, and $r_{MEED}$ for 28 six-metal carbide systems arranged in ascending order of $r_{MEED}$. The experimentally synthesized single-phase solid solutions are marked as (s).

| 6-metal carbide | $\Delta H_r$ (eV/atom) | $\langle\|\Delta E_D\|\rangle$ (eV) | $r_{MEED}$ | 6-metal carbide | $\Delta H_r$ (eV/atom) | $\langle\|\Delta E_D\|\rangle$ (eV) | $r_{MEED}$ | 6-metal carbide | $\Delta H_r$ (eV/atom) | $\langle\|\Delta E_D\|\rangle$ (eV) | $r_{MEED}$ |
|---|---|---|---|---|---|---|---|---|---|---|---|
| HfNbTaTiVZrC$_6$ (s) [18] | 0.25 | 0.83 | 0.75 | HfTaTiVWZrC$_6$ | 0.39 | 1.55 | 1.31 | HfMoNbTaWZrC$_6$ | 0.54 | 1.66 | 1.56 |
| MoNbTaTiVZrC$_6$ | 0.34 | 1.26 | 1.09 | HfNbTaTiVWC$_6$ | 0.44 | 1.45 | 1.32 | HfMoNbTaTiWC$_6$ | 0.54 | 1.66 | 1.57 |
| HfMoNbTaTiZrC$_6$ | 0.35 | 1.31 | 1.12 | HfNbTaVWZrC$_6$ | 0.44 | 1.48 | 1.33 | HfMoTiVWZrC$_6$ | 0.52 | 1.81 | 1.60 |
| HfMoNbTiVZrC$_6$ | 0.37 | 1.38 | 1.19 | MoNbTaTiVWC$_6$ | 0.51 | 1.45 | 1.42 | HfMoTaTiVWC$_6$ | 0.56 | 1.69 | 1.61 |
| HfMoTaTiVZrC$_6$ | 0.37 | 1.40 | 1.19 | MoNbTaTiWZrC$_6$ | 0.47 | 1.62 | 1.44 | HfMoNbTiVWC$_6$ | 0.57 | 1.68 | 1.61 |
| NbTaTiVWZrC$_6$ | 0.37 | 1.42 | 1.21 | MoTaTiVWZrC$_6$ | 0.49 | 1.66 | 1.49 | HfMoTaVWZrC$_6$ | 0.56 | 1.72 | 1.62 |
| HfMoNbTaTiVC$_6$ | 0.42 | 1.28 | 1.21 | MoNbTiVWZrC$_6$ | 0.49 | 1.64 | 1.49 | HfMoNbVWZrC$_6$ | 0.56 | 1.72 | 1.62 |
| HfMoNbTaVZrC$_6$ | 0.41 | 1.33 | 1.22 | MoNbTaVWZrC$_6$ | 0.53 | 1.51 | 1.49 | HfMoNbTaVWC$_6$ | 0.61 | 1.56 | 1.62 |
| HfNbTaTiWZrC$_6$ | 0.37 | 1.47 | 1.24 | HfMoTaTiWZrC$_6$ | 0.49 | 1.77 | 1.55 | | | | |
| HfNbTiVWZrC$_6$ | 0.40 | 1.53 | 1.30 | HfMoNbTiWZrC$_6$ | 0.50 | 1.76 | 1.55 | | | | |

**Table S5.** DFT-calculated formation enthalpies ($H_f$) and defect formation energies ($E_D$) of 2D sulfide systems.

| STABLEST BINARIES | | 2H BINARIES | | | | | | | | | |
|---|---|---|---|---|---|---|---|---|---|---|---|
| formula (space group) | (a) $H_f$ (eV/atom) | Formula (#187) | (b) $H_f$ (eV/atom) | (c) $E_D$ of substitutional defects (eV) | | | | | | | |
| | | | | Hf | Mo | Nb | Ta | Ti | V | W | Zr |
| $HfS_2$ (#164) | -1.58 | $HfS_2$ | -1.36 | - | 2.71 | 1.38 | 1.37 | 0.79 | 2.18 | 2.95 | 0.00 |
| $MoS_2$ (#187) | -0.89 | $MoS_2$ | -0.89 | -0.17 | - | -0.24 | -0.20 | 0.00 | 0.32 | 0.19 | -0.16 |
| $NbS_2$ (#187) | -1.05 | $NbS_2$ | -1.05 | -0.58 | 0.55 | - | 0.01 | -0.12 | 0.63 | 0.76 | -0.57 |
| $TaS_2$ (#187) | -1.05 | $TaS_2$ | -1.05 | -0.67 | 0.50 | -0.01 | - | -0.22 | 0.57 | 0.75 | -0.65 |
| $TiS_2$ (#164) | -1.29 | $TiS_2$ | -1.14 | -0.51 | 1.58 | 0.50 | 0.44 | - | 1.12 | 1.73 | -0.47 |
| $VS_2$ (#187) | -0.87 | $VS_2$ | -0.87 | -0.78 | 0.18 | -0.39 | -0.41 | -0.59 | - | 0.38 | -0.74 |
| $WS_2$ (#187) | -0.82 | $WS_2$ | -0.82 | -0.40 | -0.23 | -0.49 | -0.43 | -0.26 | 0.07 | - | -0.39 |
| $ZrS_2$ (#164) | -1.55 | $ZrS_2$ | -1.36 | 0.00 | 2.67 | 1.35 | 1.32 | 0.82 | 2.20 | 2.87 | - |

**Table S6.** Calculated values of enthalpy-component ($\Delta H_r$), entropy-component ($\langle|\Delta E_D|\rangle$) of MEED, and $r_{MEED}$ for 70 four-metal sulfide systems arranged in ascending order of $r_{MEED}$. The experimentally synthesized single-phase solid solutions are marked as (s).

| 4-metal sulfide | $\Delta H_r$ (eV/atom) | $\langle\lvert\Delta E_D\rvert\rangle$ (eV) | $r_{MEED}$ | 4-metal sulfide | $\Delta H_r$ (eV/atom) | $\langle\lvert\Delta E_D\rvert\rangle$ (eV) | $r_{MEED}$ | 4-metal sulfide | $\Delta H_r$ (eV/atom) | $\langle\lvert\Delta E_D\rvert\rangle$ (eV) | $r_{MEED}$ |
|---|---|---|---|---|---|---|---|---|---|---|---|
| $MoNbTaVS_8$ (s) [19] | 0.09 | 0.46 | 0.46 | $MoTaTiZrS_8$ | 0.44 | 1.11 | 1.34 | $HfMoNbVS_8$ | 0.54 | 1.25 | 1.55 |
| $MoTaVWS_8$ | 0.14 | 0.46 | 0.51 | $NbTaVZrS_8$ | 0.47 | 1.09 | 1.35 | $HfMoTaVS_8$ | 0.54 | 1.25 | 1.56 |
| $MoNbTaWS_8$ (s) [19] | 0.10 | 0.53 | 0.52 | $HfNbTiVS_8$ | 0.47 | 1.10 | 1.37 | $HfNbWZrS_8$ | 0.43 | 1.44 | 1.57 |
| $MoNbVWS_8$ (s) [19] | 0.15 | 0.49 | 0.53 | $HfTiVZrS_8$ | 0.39 | 1.23 | 1.37 | $HfMoTiVS_8$ | 0.51 | 1.32 | 1.57 |
| $NbTaVWS_8$ | 0.10 | 0.57 | 0.57 | $HfTaTiVS_8$ | 0.47 | 1.10 | 1.37 | $MoTaWZrS_8$ | 0.52 | 1.30 | 1.58 |
| $MoNbTaTiS_8$ | 0.25 | 0.55 | 0.71 | $HfMoNbTiS_8$ | 0.47 | 1.12 | 1.38 | $MoNbWZrS_8$ | 0.52 | 1.30 | 1.58 |
| $NbTaTiVS_8$ | 0.26 | 0.61 | 0.75 | $HfMoTaTiS_8$ | 0.47 | 1.13 | 1.39 | $HfTaWZrS_8$ | 0.43 | 1.46 | 1.58 |
| $NbTaTiWS_8$ | 0.27 | 0.72 | 0.85 | $HfMoTiZrS_8$ | 0.39 | 1.26 | 1.39 | $NbVWZrS_8$ | 0.53 | 1.31 | 1.59 |
| $MoTaTiVS_8$ | 0.30 | 0.73 | 0.89 | $HfMoNbTaS_8$ | 0.49 | 1.09 | 1.39 | $TaVWZrS_8$ | 0.53 | 1.31 | 1.59 |
| $MoNbTiVS_8$ | 0.30 | 0.74 | 0.89 | $HfNbTaVS_8$ | 0.50 | 1.10 | 1.40 | $MoTiWZrS_8$ | 0.50 | 1.37 | 1.60 |
| $MoTaTiWS_8$ | 0.31 | 0.80 | 0.96 | $NbTiWZrS_8$ | 0.46 | 1.22 | 1.43 | $TiVWZrS_8$ | 0.50 | 1.38 | 1.61 |
| $MoNbTiWS_8$ | 0.31 | 0.82 | 0.97 | $TaTiWZrS_8$ | 0.46 | 1.22 | 1.44 | $HfMoNbWS_8$ | 0.55 | 1.33 | 1.63 |
| $TaTiVWS_8$ | 0.32 | 0.82 | 0.97 | $NbTaWZrS_8$ | 0.48 | 1.18 | 1.44 | $HfMoTaWS_8$ | 0.55 | 1.33 | 1.63 |
| $NbTiVWS_8$ | 0.32 | 0.83 | 0.98 | $HfNbVZrS_8$ | 0.42 | 1.29 | 1.44 | $HfNbVWS_8$ | 0.55 | 1.33 | 1.64 |
| $MoTiVWS_8$ | 0.36 | 0.82 | 1.03 | $HfMoNbZrS_8$ | 0.41 | 1.31 | 1.45 | $HfMoVZrS_8$ | 0.46 | 1.49 | 1.64 |
| $HfNbTiZrS_8$ | 0.35 | 0.85 | 1.04 | $HfTaVZrS_8$ | 0.42 | 1.30 | 1.45 | $HfTaVWS_8$ | 0.56 | 1.34 | 1.64 |
| $HfTaTiZrS_8$ | 0.35 | 0.88 | 1.06 | $HfMoTaZrS_8$ | 0.41 | 1.33 | 1.46 | $HfMoTiWS_8$ | 0.53 | 1.40 | 1.65 |
| $NbTaTiZrS_8$ | 0.40 | 0.81 | 1.08 | $HfNbTiWS_8$ | 0.49 | 1.24 | 1.49 | $MoVWZrS_8$ | 0.57 | 1.33 | 1.65 |
| $HfNbTaTiS_8$ | 0.43 | 0.83 | 1.13 | $HfTaTiWS_8$ | 0.49 | 1.25 | 1.49 | $HfTiVWS_8$ | 0.53 | 1.40 | 1.66 |
| $HfNbTaZrS_8$ | 0.37 | 0.97 | 1.15 | $HfNbTaWS_8$ | 0.51 | 1.21 | 1.50 | $HfMoWZrS_8$ | 0.47 | 1.55 | 1.70 |
| $NbTiVZrS_8$ | 0.44 | 1.09 | 1.32 | $MoNbVZrS_8$ | 0.51 | 1.23 | 1.51 | $HfMoVWS_8$ | 0.60 | 1.35 | 1.70 |
| $TaTiVZrS_8$ | 0.45 | 1.09 | 1.33 | $MoTaVZrS_8$ | 0.51 | 1.23 | 1.51 | $HfVWZrS_8$ | 0.48 | 1.60 | 1.74 |
| $MoNbTiZrS_8$ | 0.44 | 1.10 | 1.33 | $HfTiWZrS_8$ | 0.41 | 1.40 | 1.51 | | | | |
| $MoNbTaZrS_8$ | 0.46 | 1.07 | 1.34 | $MoTiVZrS_8$ | 0.49 | 1.30 | 1.53 | | | | |

**Table S7.** Calculated values of enthalpy-component ($\Delta H_r$), entropy-component ($\langle|\Delta E_D|\rangle$) of MEED, and $r_{MEED}$ for 56 five-metal sulfide systems arranged in ascending order of $r_{MEED}$. The experimentally synthesized single-phase solid solutions are marked as (s).

| **5-metal sulfide** | **$\Delta H_r$ (eV/atom)** | **$\langle\|\Delta E_D\|\rangle$ (eV)** | **$r_{MEED}$** | **5-metal sulfide** | **$\Delta H_r$ (eV/atom)** | **$\langle\|\Delta E_D\|\rangle$ (eV)** | **$r_{MEED}$** | **5-metal sulfide** | **$\Delta H_r$ (eV/atom)** | **$\langle\|\Delta E_D\|\rangle$ (eV)** | **$r_{MEED}$** |
|---|---|---|---|---|---|---|---|---|---|---|---|
| $MoNbTaVWS_{10}$ (s) [19] | 0.12 | 0.49 | 0.51 | $MoNbTaVZrS_{10}$ | 0.51 | 1.01 | 1.36 | $HfMoTaTiWS_{10}$ | 0.53 | 1.16 | 1.49 |
| $MoNbTaTiVS_{10}$ | 0.29 | 0.61 | 0.79 | $MoNbTiVZrS_{10}$ | 0.49 | 1.08 | 1.38 | $HfMoTiVZrS_{10}$ | 0.45 | 1.30 | 1.49 |
| $MoNbTaTiWS_{10}$ | 0.30 | 0.67 | 0.85 | $MoTaTiVZrS_{10}$ | 0.49 | 1.08 | 1.38 | $HfTaTiVWS_{10}$ | 0.53 | 1.16 | 1.49 |
| $NbTaTiVWS_{10}$ | 0.30 | 0.70 | 0.87 | $HfNbTiWZrS_{10}$ | 0.43 | 1.21 | 1.40 | $MoTaVWZrS_{10}$ | 0.55 | 1.13 | 1.50 |
| $MoTaTiVWS_{10}$ | 0.33 | 0.72 | 0.93 | $HfTaTiWZrS_{10}$ | 0.43 | 1.22 | 1.41 | $MoNbVWZrS_{10}$ | 0.55 | 1.14 | 1.51 |
| $MoNbTiVWS_{10}$ | 0.33 | 0.73 | 0.93 | $HfMoNbTaVS_{10}$ | 0.54 | 1.03 | 1.41 | $HfMoNbVZrS_{10}$ | 0.47 | 1.29 | 1.51 |
| $HfNbTaTiZrS_{10}$ | 0.39 | 0.85 | 1.08 | $MoNbTaWZrS_{10}$ | 0.52 | 1.07 | 1.41 | $HfMoTaVZrS_{10}$ | 0.47 | 1.30 | 1.51 |
| $MoNbTaTiZrS_{10}$ | 0.45 | 0.92 | 1.22 | $HfMoNbTiVS_{10}$ | 0.52 | 1.09 | 1.42 | $MoTiVWZrS_{10}$ | 0.54 | 1.23 | 1.54 |
| $NbTaTiVZrS_{10}$ | 0.46 | 0.93 | 1.23 | $NbTaVWZrS_{10}$ | 0.52 | 1.08 | 1.43 | $HfMoTiWZrS_{10}$ | 0.47 | 1.37 | 1.55 |
| $HfMoNbTaTiS_{10}$ | 0.48 | 0.93 | 1.27 | $HfMoTaTiVS_{10}$ | 0.52 | 1.09 | 1.43 | $HfMoTaVWS_{10}$ | 0.58 | 1.15 | 1.56 |
| $HfNbTaTiVS_{10}$ | 0.48 | 0.94 | 1.28 | $HfNbTaWZrS_{10}$ | 0.45 | 1.23 | 1.43 | $HfMoNbVWS_{10}$ | 0.58 | 1.15 | 1.56 |
| $HfNbTiVZrS_{10}$ | 0.42 | 1.09 | 1.30 | $MoNbTiWZrS_{10}$ | 0.50 | 1.14 | 1.44 | $HfTiVWZrS_{10}$ | 0.47 | 1.38 | 1.56 |
| $HfTaTiVZrS_{10}$ | 0.42 | 1.10 | 1.31 | $MoTaTiWZrS_{10}$ | 0.50 | 1.14 | 1.44 | $HfMoNbWZrS_{10}$ | 0.48 | 1.36 | 1.57 |
| $NbTaTiWZrS_{10}$ | 0.47 | 1.02 | 1.31 | $NbTiVWZrS_{10}$ | 0.50 | 1.14 | 1.44 | $HfMoTaWZrS_{10}$ | 0.48 | 1.37 | 1.58 |
| $HfMoNbTiZrS_{10}$ | 0.42 | 1.11 | 1.31 | $TaTiVWZrS_{10}$ | 0.50 | 1.15 | 1.44 | $HfNbVWZrS_{10}$ | 0.49 | 1.37 | 1.58 |
| $HfMoTaTiZrS_{10}$ | 0.42 | 1.12 | 1.32 | $HfMoNbTaWS_{10}$ | 0.55 | 1.09 | 1.47 | $HfTaVWZrS_{10}$ | 0.49 | 1.38 | 1.58 |
| $HfNbTaVZrS_{10}$ | 0.44 | 1.12 | 1.35 | $HfNbTaVWS_{10}$ | 0.55 | 1.10 | 1.48 | $HfMoTiVWS_{10}$ | 0.56 | 1.25 | 1.59 |
| $HfMoNbTaZrS_{10}$ | 0.44 | 1.13 | 1.35 | $HfMoNbTiWS_{10}$ | 0.53 | 1.16 | 1.49 | $HfMoVWZrS_{10}$ | 0.52 | 1.43 | 1.67 |
| $HfNbTaTiWS_{10}$ | 0.50 | 1.04 | 1.36 | $HfNbTiVWS_{10}$ | 0.53 | 1.16 | 1.49 | | | | |

**Table S8.** Calculated values of enthalpy-component ($\Delta H_r$), entropy-component ($\langle|\Delta E_D|\rangle$) of MEED, and $r_{MEED}$ for 28 six-metal sulfide systems arranged in ascending order of $r_{MEED}$.

| **6-metal sulfide** | **$\Delta H_r$ (eV/atom)** | **$\langle\|\Delta E_D\|\rangle$ (eV)** | **$r_{MEED}$** | **6-metal sulfide** | **$\Delta H_r$ (eV/atom)** | **$\langle\|\Delta E_D\|\rangle$ (eV)** | **$r_{MEED}$** | **6-metal sulfide** | **$\Delta H_r$ (eV/atom)** | **$\langle\|\Delta E_D\|\rangle$ (eV)** | **$r_{MEED}$** |
|---|---|---|---|---|---|---|---|---|---|---|---|
| $MoNbTaTiVWS_{12}$ | 0.32 | 0.65 | 0.86 | $HfMoNbTiVZrS_{12}$ | 0.47 | 1.15 | 1.40 | $HfTaTiVWZrS_{12}$ | 0.48 | 1.22 | 1.46 |
| $HfNbTaTiVZrS_{12}$ | 0.44 | 0.99 | 1.26 | $MoNbTaVWZrS_{12}$ | 0.55 | 0.99 | 1.40 | $HfMoNbTaWZrS_{12}$ | 0.49 | 1.20 | 1.47 |
| $HfMoNbTaTiZrS_{12}$ | 0.44 | 1.00 | 1.26 | $HfMoTaTiVZrS_{12}$ | 0.47 | 1.16 | 1.40 | $HfNbTaVWZrS_{12}$ | 0.49 | 1.20 | 1.47 |
| $MoNbTaTiVZrS_{12}$ | 0.49 | 0.94 | 1.29 | $HfMoNbTaVZrS_{12}$ | 0.48 | 1.14 | 1.41 | $HfMoNbTiVWS_{12}$ | 0.56 | 1.09 | 1.48 |
| $HfNbTaTiWZrS_{12}$ | 0.45 | 1.08 | 1.33 | $MoNbTiVWZrS_{12}$ | 0.53 | 1.07 | 1.43 | $HfMoTaTiVWS_{12}$ | 0.56 | 1.09 | 1.48 |
| $HfMoNbTaTiVS_{12}$ | 0.52 | 0.95 | 1.34 | $MoTaTiVWZrS_{12}$ | 0.53 | 1.07 | 1.43 | $HfMoNbVWZrS_{12}$ | 0.52 | 1.28 | 1.56 |
| $MoNbTaTiWZrS_{12}$ | 0.50 | 0.99 | 1.34 | $HfMoNbTaVWS_{12}$ | 0.57 | 1.01 | 1.45 | $HfMoTiVWZrS_{12}$ | 0.51 | 1.31 | 1.56 |
| $NbTaTiVWZrS_{12}$ | 0.50 | 1.00 | 1.34 | $HfMoNbTiWZrS_{12}$ | 0.48 | 1.22 | 1.46 | $HfMoTaVWZrS_{12}$ | 0.52 | 1.29 | 1.56 |
| $HfMoNbTaTiWS_{12}$ | 0.53 | 1.01 | 1.39 | $HfNbTiVWZrS_{12}$ | 0.48 | 1.21 | 1.46 | | | | |
| $HfNbTaTiVWS_{12}$ | 0.53 | 1.01 | 1.39 | $HfMoTaTiWZrS_{12}$ | 0.48 | 1.22 | 1.46 | | | | |

**Note S1. Methods**

First-principles density functional theory (DFT) calculations were performed using the projector-augmented wave (PAW) method [20,21] as implemented in the Vienna Ab initio Simulation Package (VASP). [22,23] A kinetic energy cutoff of 500 eV for carbides and 400 eV for sulfides were used for the plane-wave expansion. Total energies were calculated using the Perdew-Burke-Ernzerhof exchange-correlation functional [24] within spin-polarized approaches. All 2D calculations were performed on single $MoS_2$ monolayer (2H phase) slab structure model. A vacuum spacing of 18Å was used to reduce the interaction between image planes along the direction normal to the surface.

DFT simulations on two-metal (ternary) and five-metal disordered alloys were performed using special quasi-random structures (SQS) [25] generated by the Alloy Theoretic Automated Toolkit (ATAT). [26] POSCAR of each representative SQS structure is listed in Supplemental material S1-S2. Various supercell sizes (including binaries and point-defect supercells) and their corresponding k-points are listed in Supplementary Table S9. All atoms were fully relaxed until their atomic forces were less than 0.001 eV/Å for binaries, 0.01 eV/Å for ternaries and point-defect supercells, and 0.02 eV/Å for five-metal SQS supercells.

**Table S9.** Supercell sizes and corresponding k-points used in DFT calculations to estimate MEED parameters.

| **Calculations** | **carbide supercell size** | **k-points** | **2D sulfide supercell size** | **k-points** |
|---|---|---|---|---|
| heat of formation of binaries ($H_f^{bi}$) | 1x1x1 | 12x12x12 | 1x1x1 | 16x16x1 |
| defect formation energy (point defect) | 2x2x2 | 6x6x6 | 4x2√3x1 | 4x4x1 |
| heat of formation of ternaries ($H_f^{tr}$) | 2x2x1 (Fig. S1a) | 4x4x8 | 4x4x1 (Fig. S1c) | 4x4x1 |
| heat of formation of five-metal compounds ($H_f^{HEM}$) | 2x2x5 (Fig. S1b) | 2x2x1 | 5x13x1 (Fig. S1d) | 1x1x1 |

The heat of formation ($H_f$) of a binary compound $M_mX_n$ ($m$ and $n$ are integers) is calculated by using the formula: $H_f = \{E_{M_mX_n} - m\mu_M - n\mu_X\}/(m+n)$, where $E_{M_mX_n}$ is the total energy (per formula unit) of the $M_mX_n$ compound, and $\mu_M$ and $\mu_X$ are the chemical potentials of M and X in their elemental metal or gas phases, respectively.

The defect formation energy ($E_D$) for a substitutional point defect is calculated from formula $E_D = E_D^t + \mu_H - \mu_D - E_H^t$, where $E_D^t$ and $E_H^t$ are the total energies of the supercells containing single defect and zero defect (host binary), respectively. $\mu_H$ and $\mu_D$ are the chemical potentials of the host and defect metal atoms estimated from the total energy (per atom) of pure metals.

## Note S2. POSCARs for two-metal (ternary) and five-metal carbide SQS structures

```
######################################################################
# POSCAR for 2-metal (ternary) carbides (e.g., HfNbC2)
######################################################################
POSCAR
  4.65000000000000
    1.9618687016406757   -0.0018793713179200    0.0000000000000000
   -0.0018767120378116    1.9629157375110036    0.0000000000000000
    0.0000000000000000    0.0000000000000000    0.9808912787755264
  Nb   Hf   C
    8    8   16
Direct
  0.4983674605941685  0.5022897074579241  0.0000000000000000
  0.7504858879427695  0.5026465152155323  0.5000000000000000
  0.7571070809815426  0.7509300812475476  0.0000000000000000
  0.0010631154695465  0.7524370730829532  0.5000000000000000
  0.0032657938046681 -0.0041603307992695  0.0000000000000000
  0.2427503523233111  0.7569016815256793  0.0000000000000000
  0.2474108549700716 -0.0017322124828787  0.5000000000000000
  0.2481059066311778  0.2425457470783275  0.0000000000000000
  0.4991709093257347  0.7493582972737912  0.5000000000000000
  0.4982068774044995 -0.0017749909579471  0.0000000000000000
  0.4987368097995694  0.2507432475422610  0.5000000000000000
  0.7516644329750843 -0.0015904672914623  0.5000000000000000
  0.0010348940613976  0.5017394816770361  0.0000000000000000
  0.2500612677996714  0.5009552847094829  0.5000000000000000
  0.7506160176661286  0.2503184653938521  0.0000000000000000
  0.0011660959185086  0.2483252817048556  0.5000000000000000
  0.2497543629272136  0.2483273139221990  0.5000000000000000
  0.2529217190587430  0.5002142679484204  0.0000000000000000
  0.4973810788486003  0.2533150406166836  0.0000000000000000
  0.2481632475937803  0.7515917819087502  0.5000000000000000
  0.2482648530388536 -0.0002182831924989  0.0000000000000000
  0.5010154702174989  0.5011394518612143  0.5000000000000000
  0.5008473250466295  0.7462552051991759  0.0000000000000000
  0.7499388274338703  0.2533301777832149  0.5000000000000000
  0.7483395149794312  0.5029972186287115  0.0000000000000000
  0.0030227883690383  0.2490744379440823  0.0000000000000000
  0.4973759203441185 -0.0011059537392171  0.5000000000000000
  0.7517078276624591  0.7475702221440433  0.5000000000000000
  0.7521344757907690 -0.0032701263523003  0.0000000000000000
 -0.0019384965208569  0.5024292994892625  0.5000000000000000
 -0.0006276808083265  0.7507541794659230  0.0000000000000000
  0.0024850083503286 -0.0023370960053494  0.5000000000000000
```

```
######################################################################
# POSCAR for 5-metal carbides (e.g., HfNbTaTiZrC5)
######################################################################
POSCAR
  4.85000000000000
    1.8663780020618923   -0.0003871838693130    0.0002676521765037
   -0.0003863911643682    1.8677739435504950   -0.0004378963145989
    0.0006634813296177   -0.0010654835687638    4.6939718353896094
  Hf   Nb   Ta   Ti   Zr   C
   16    16    16    16    16    80
Direct
  0.4986520546248604  0.4995182123393756  0.0017239871361365
  0.4983515252978141  0.5010715446531151  0.8012096753776299
  0.4983973003321845  0.0003304690291614  0.0007612865747876
  0.7494389043367570  0.7488984492045532  0.0003390320491806
  0.7496112493928414  0.0022998768422395  0.1017245696002119
  0.4982102230831535  0.2526956453222567  0.9005123182407131
  0.7512589161027294  0.2512609357598347  0.0012984089484496
  0.2502654536014949  0.7480961121479123  0.8004513417967350
  0.2486906137341535  0.2519116655360247  0.0008498876746260
  0.4980263687277940  0.4960368428824389  0.2002490146130175
  0.4991514120253662  0.2470039496287821  0.4977622875678480
  0.7499571621528683  0.2495352962009367  0.6000829248057075
  0.0002066051251555  0.4985231956678153  0.3972996562282757
  0.0031759664962641  0.7509256365418955  0.5000637106023059
  0.2477360275401505  0.4957670323929488  0.2975892850141467
  0.2503941077106016  0.2491543107066517  0.5993917699668583
  0.4946606650842225  0.7505878190978077  0.1006227530960778
  0.4899265104515568  0.7465986161308609  0.9024994098119161
  0.0023694977819137  0.5042070380703688  0.9983728416223080
  0.2484081663760239  0.4973340424933899  0.1018117072835574
  0.4979710515681076  0.5065247306175060  0.5988431045515514
  0.0108640168366088  0.7442538756399055  0.9018143071707776
  0.0020934027347942  0.2506504649736940  0.1039167296586495
  0.2494543567333381  0.7471760312823288  0.0001504355573626
  0.4970240735556646  0.7530091680654492  0.5001582216002164
  0.4949645464804096  0.9936275632884585  0.6010423918397600
  0.7502161884874874  0.2505475253461454  0.8013527991670519
  0.7547732055346473  0.0083881193494510  0.2981551461229767
  0.9993716455710401  0.7608252813333284  0.2976588095446193
  0.9977993653661602  0.2423162455120073  0.2970686881328641
  0.2484422151135832  0.7564796497351768  0.2000497075995143
  0.2507000006730104  0.2454501676596954  0.3969361508957391
  0.7513386829194427  0.4977161248715823  0.1024188229130576
  0.4975802329196782  0.9975096707955782  0.7996716974745539
  0.7453771521054952  0.9974142022960104  0.8999323823993824
  0.0017791733653816  0.5013916342068079  0.8031678459323440
  0.0012010509458614  0.9970711443652130  0.0000662194885170
  0.4997657111755292  0.4965798162936183  0.3981647205079764
  0.7448644338917985  0.9972707702825395  0.7011641737283341
  0.2529407813773757  0.9974574861642913  0.8993917847628524
  0.7524380739281702  0.4904822817505311  0.2978182528399374
  0.0057154282919073  0.9973069169742479  0.5999141034601764
  0.0014157647666472  0.2490102785918569  0.7005985249651395
  0.2545161692110179  0.9966162707697236  0.7006177505703537
  0.7508114312540846  0.7582152496205729  0.2004447024810851
  0.2519644491575347  0.7547457850785060  0.3990375712498776
  0.7518629381376486  0.2436795737452368  0.2009796580501028
  0.2479267556128512  0.0054244630185338  0.2982895632128436
  0.4979456352304039  0.2496596493451174  0.1027143710999433
  0.7488660288318512  0.5033199413870920  0.9013037493533315
  0.4975600190681718  0.7499113182288815  0.6984758312029736
  0.2501142450277890  0.5045421180667041  0.9016559431075064
  0.0020814491842978  0.5023869239431838  0.6003335629124941
  0.0028733676647534  0.7511625069136835  0.6998549969320529
  0.4999891159042289  0.0040890396194238  0.3977582631564007
  0.2496302929906520  0.7509592149601712  0.6000601873016748
  0.2496779584399259  0.2511369741695965  0.8010883015784955
  0.4978940823875830  0.0043188620541985  0.2005151734781908
  0.5023433395070509  0.2458666799068310  0.2983552954109905
```

| | | |
|---|---|---|
| 0.7490187179982409 | 0.2474846710732416 | 0.3959950498915717 |
| 0.0020293732135963 | 0.4948599633511298 | 0.2001258378247632 |
| 0.2529097090783543 | 0.9991856464318405 | 0.4993216526959092 |
| 0.0021516195508207 | 0.0039117407764863 | 0.2004504550750898 |
| 0.2471407749169454 | 0.2449004157954144 | 0.2012888149498940 |
| 0.0028115326557899 | 0.7489182698876673 | 0.1009785938887586 |
| 0.7517818280126184 | 0.5020007203663460 | 0.7005285970895381 |
| 0.7498367124873224 | 0.7488471296201574 | 0.8015394897466207 |
| 0.2497918774256828 | 0.0033587761248738 | 0.1022689771683611 |
| 0.4988336126858391 | 0.2501589264302363 | 0.7004143501012593 |
| 0.7495666976265765 | 0.4991765606275176 | 0.4982534707441769 |
| 0.7513160782910661 | 0.7505035449550166 | 0.6005270110905545 |
| 0.0022084381499727 | 0.9979015930114244 | 0.8007896498870287 |
| 0.0015689532743340 | 0.2543522383059525 | 0.9003377294751745 |
| 0.2483045105988306 | 0.5026923409166086 | 0.6999350241488009 |
| 0.5006053458803699 | 0.7545580167556079 | 0.2987233924157573 |
| 0.7482387217204115 | 0.7523705080207264 | 0.3975208922017861 |
| 0.7471297955378143 | 0.0007251723430750 | 0.4988302029838891 |
| 0.2503959780353815 | 0.5011195718855761 | 0.4991539870428947 |
| 0.0006816747762769 | 0.0023970953134410 | 0.3985616100931294 |
| 0.0008567239787134 | 0.2491046366883204 | 0.4991571813621767 |
| 0.2503870937280953 | 0.2463726817376082 | 0.8990614731230204 |
| 0.2471022362076516 | 0.5015341064282913 | 0.0001964670714260 |
| 0.5000239614975621 | 0.2499437399387313 | 0.0024041014434981 |
| 0.2447531603969255 | 0.2447771395085111 | 0.7011993275067592 |
| 0.2459063775010780 | 0.4958446811186841 | 0.8028467096510208 |
| 0.2495798583266209 | 0.7490745562804650 | 0.9010168823378075 |
| 0.2459416449117657 | 0.9981417386810491 | 0.9985187836499531 |
| 0.4989282282583201 | 0.2463730518582782 | 0.8013263782045821 |
| 0.5016031439366254 | 0.5032425022978261 | 0.9009491143817072 |
| 0.4994406924487302 | 0.7493715307899352 | 0.0005160794986931 |
| 0.7491714997114222 | 0.2480582857462699 | 0.9001150274754981 |
| 0.7535422819893471 | 0.5005231824483325 | 0.0013230613199974 |
| 0.0001759043512187 | 0.2475829138517808 | 0.0023533048258999 |
| 0.2520324829285570 | 0.2437122974933274 | 0.4979558465015675 |
| 0.2487836055903733 | 0.5049946169164140 | 0.5998967864329248 |
| 0.2494503536935440 | 0.7554282180800515 | 0.6984623299352117 |
| 0.2528068307662759 | 0.0025499182103748 | 0.8003138741417622 |
| 0.4992593232217970 | 0.2487448467888816 | 0.5990431365139138 |
| 0.4994293995568438 | 0.5063579477383432 | 0.6994479493113644 |
| 0.4983415750363507 | 0.7541725155125663 | 0.8002070420456289 |
| 0.5011700441936109 | 0.9989516097034664 | 0.8991527868157408 |
| 0.7548033327881422 | 0.2452329380024954 | 0.7009833826451883 |
| 0.7535191280729282 | 0.4964483516207530 | 0.8035192046661102 |
| 0.7508401111391906 | 0.7502546315358738 | 0.9016977049273009 |
| 0.7533511486358254 | 0.9995481278911520 | 0.9992129157412803 |
| 0.0015800535817676 | 0.2522779670036120 | 0.7990544973434387 |
| 0.9986947206418364 | 0.5024398920010005 | 0.9001947050567451 |
| 0.0005222014635743 | 0.7521675945319709 | 0.9993641021691896 |
| 0.2544377873579577 | 0.2472985790136681 | 0.2982117142736358 |
| 0.2539444554352683 | 0.5019435683375536 | 0.3973549975625733 |
| 0.2537764537318648 | 0.7563902055370201 | 0.4992132628600701 |
| 0.2485727183912557 | 0.9945477522815144 | 0.5996352126681890 |
| 0.5009061931311155 | 0.2488541455585605 | 0.3961072364419364 |
| 0.4991380789683983 | 0.5014158647316264 | 0.4978873557567784 |
| 0.4933778885617228 | 0.7505955665308970 | 0.6005571702745172 |
| 0.5002172350730495 | 0.9945931956367369 | 0.7006263630111167 |
| 0.7478537445877292 | 0.2496561089305105 | 0.4982719626578297 |
| 0.7508561851389446 | 0.4995259290217140 | 0.6001141669909213 |
| 0.7506265538392469 | 0.7562030918981441 | 0.7002354774961178 |
| 0.7474362804035950 | 0.0022356181938119 | 0.8009854260221850 |
| 0.0009949129900077 | 0.2482780817240030 | 0.6020723135718146 |
| 0.0005200376516322 | 0.4993741954796055 | 0.7006031849391879 |
| 0.0014994737055318 | 0.7450913418398523 | 0.7996660032139751 |
| 0.9989472359994603 | 0.9977025195700814 | 0.9026571514482647 |
| 0.2508815066615004 | 0.2546819566378281 | 0.1026002350032377 |
| 0.2445206866797827 | 0.4994274413470993 | 0.1986273940707892 |
| 0.2472266504995752 | 0.7526520906484209 | 0.2995608366764905 |
| 0.2565817526230916 | 0.9995693761794346 | 0.3988512330691251 |
| 0.5012877000402557 | 0.2448388345180935 | 0.2004185166843316 |
| 0.5026287876984475 | 0.4951079075739484 | 0.3002545805988345 |
| 0.4953469481171870 | 0.7510771804497248 | 0.3999648466890142 |
| 0.4927296708138239 | 0.9973951042598099 | 0.4983777455794292 |
| 0.7449978747323951 | 0.2521468887632396 | 0.2993058914516974 |
| 0.7459837651986588 | 0.4950494533546376 | 0.3957976213837276 |
| 0.7458654575455712 | 0.7505847390170275 | 0.4994206343150799 |
| 0.7527188795809791 | 0.0002786834186651 | 0.6022481966844236 |
| 0.0000272090193855 | 0.2512502405448919 | 0.3969231821508281 |
| 0.0008020122353345 | 0.5016421383885958 | 0.5002638511639221 |
| 0.0070787022982066 | 0.7505038671818274 | 0.6019601865224770 |
| 0.0000835039491477 | 0.9990788994640025 | 0.6986326023057625 |
| 0.2529545438581685 | 0.7467232230548601 | 0.1014220871060193 |
| 0.2514343175952264 | 0.0028479020080567 | 0.2021508086444758 |
| 0.5014936044218020 | 0.4990494132675724 | 0.1017609074533725 |
| 0.5026433909474174 | 0.7536896391870888 | 0.1985640322403652 |
| 0.4976869191586771 | 0.0080706189421689 | 0.2991302133403873 |
| 0.7487549724813934 | 0.2536952458326639 | 0.1026332134524525 |
| 0.7541718684111308 | 0.4996195412153272 | 0.1996873610343737 |
| 0.7540309596119051 | 0.7463897655219179 | 0.2967557772549043 |
| 0.7423801333657140 | 0.0066445679176937 | 0.3971105839514413 |
| 0.0004168629180123 | 0.2513864780569506 | 0.2000505011342942 |
| 0.9962362789859196 | 0.5010248638441636 | 0.2961294077560611 |
| 0.0057943621711329 | 0.7496581949861629 | 0.3975642810211539 |
| 0.0061421103183972 | 0.9985164270727320 | 0.5014134460545648 |
| 0.5006579905362470 | 0.0004547091206484 | 0.1027607304303631 |
| 0.7472421763749498 | 0.7451183069526932 | 0.1025720322875531 |
| 0.7482514900180666 | 0.9997247340287978 | 0.2005327995833276 |
| 0.9983551800699999 | 0.4961982850231374 | 0.1023372700800092 |
| 0.9979917255060403 | 0.7497923143133468 | 0.2015975205706236 |
| 0.0034547582846850 | 0.9995339137547059 | 0.2969253884818883 |
| 0.9986843799042712 | 0.0039717565987353 | 0.1009043305475042 |

## Note S3. POSCARs for two-metal (ternary) and five-metal 2D-sulfide SQS structures

```
######################################################################
# POSCAR for 2-metal (ternary) sulfides (e.g., MoNbS4)
######################################################################
POSCAR
   1.00000000000000
    13.0967104774900047    0.0000000000000000    0.0000000000000000
    -6.5483559633493806   11.3460393231667265    0.0000000000000000
     0.0000000000000000    0.0000000000000000   18.0000000000000000
   Nb   Mo   S
     8     8    32
Direct
  0.9995989468475486  0.0063671416949589  0.1706611219999985
  0.2570419098293542  0.0039000580998731  0.1706611219999985
  0.5039951357593750  0.2522022181469765  0.1706611219999985
  0.9964903714477060  0.4925495905375001  0.1706611219999985
  0.9931746988987840  0.7455214010112456  0.1706611219999985
  0.2563913196443082  0.7525198715752026  0.1706611219999985
  0.5020021041474010  0.7535459333872438  0.1706611219999985
  0.7496549427976404  0.7478481296796886  0.1706611219999985
  0.5019297805216567  0.0011380902219571  0.1706611219999985
  0.7492168572622120  0.0038978880888365  0.1706611219999985
  0.9985194656207597  0.2484127675957666  0.1706611219999985
  0.2488852751608590  0.2530810095366292  0.1706611219999985
  0.7500758950652582  0.2482849821892188  0.1706611219999985
  0.2502238085948818  0.4971774247248959  0.1706611219999985
  0.4960398343453463  0.4958498756115119  0.1706611219999985
  0.7496692754452070  0.4991749001319832  0.1706611219999985
  0.1693888865246009  0.0881877849881505  0.0848579937400160
  0.4215657310981697  0.0839820414001551  0.0833642065853795
  0.6663543740774998  0.0844970353321983  0.0853458055109542
  0.9132325014016871  0.0861238272360652  0.0840598511431097
  0.1643285416407423  0.3325518749999787  0.0856301127800947
  0.4139633744234956  0.3331292692826935  0.0850860524963579
  0.6713760623161704  0.3346621827051663  0.0846485154396035
  0.9136760944888564  0.3269794672789104  0.0842548439181101
  0.1682613679647957  0.5808584354467854  0.0850539565582196
  0.4154600934532553  0.5801697382758988  0.0855030199489093
  0.6626959048180439  0.5779294712773790  0.0852500230297224
  0.9130366900797924  0.5791484952464074  0.0837698191640612
  0.1643429561849956  0.8317621314587456  0.0848782155366976
  0.4205206928394176  0.8383617054149965  0.0829973567600177
  0.6693321875058587  0.8377662007606617  0.0845722000858302
  0.9176775184884320  0.8364867297790894  0.0833135710658084
  0.1693888865246009  0.0881877849881505  0.2564642272599897
  0.4215657310981697  0.0839820414001551  0.2579580144146263
  0.6663543740774998  0.0844970353321983  0.2559764154890516
  0.9132325014016871  0.0861238272360652  0.2572623698568961
  0.1643285416407423  0.3325518749999787  0.2556921082199111
  0.4139633744234956  0.3331292692826935  0.2562361685036478
  0.6713760623161704  0.3346621827051663  0.2566737055604023
  0.9136760944888564  0.3269794672789104  0.2570673770818956
  0.1682613679647957  0.5808584354467854  0.2562682644417862
  0.4154600934532553  0.5801697382758988  0.2558192010510965
  0.6626959048180439  0.5779294712773790  0.2560721979702834
  0.9130366900797924  0.5791484952464074  0.2575524018359445
  0.1643429561849956  0.8317621314587456  0.2564440054633081
  0.4205206928394176  0.8383617054149965  0.2583248642399880
  0.6693321875058587  0.8377662007606617  0.2567500209141755
  0.9176775184884320  0.8364867297790894  0.2580086499341974
```

```
######################################################################
# POSCAR for 5-metal sulfides (e.g., MoNbTaVWS10)
######################################################################
POSCAR
   1.00000000000000
    16.2805583454967930    0.0000000000000000    0.0000000000000000
   -21.1647180006915683   36.5891805909262473    0.0000000000000000
     0.0000000000000000    0.0000000000000000   18.0000000000000000
   Mo   Nb   Ta   V    W    S
    13    13    13    13    13   130
Direct
  0.0015071418503041  0.0023754110552474  0.4827077690000010
  0.2027553744766308  0.1554660061170949  0.4827077690000010
  0.4001010110358010  0.1554379096831511  0.4827077690000010
  0.6025759304226455  0.1552262746989825  0.4827077690000010
  0.8004254241573889  0.1554390080298660  0.4827077690000010
  0.2021132882432042  0.2310758945973248  0.4827077690000010
  0.7985264546351374  0.2302522986509246  0.4827077690000010
  0.7981379122385661  0.4600450104144826  0.4827077690000010
  0.9998838412006918  0.5364884529832992  0.4827077690000010
  0.6015581881779823  0.6152577958796215  0.4827077690000010
  0.4002218518569052  0.8448068042126096  0.4827077690000010
  0.1987855235871123  0.9227316977131679  0.4827077690000010
  0.8012648123821293  0.9243759398617186  0.4827077690000010
  0.1989747695458135  0.0009060593521113  0.4827077690000010
  0.4050726941506468  0.0806913484707295  0.4827077690000010
  0.8037996840316808  0.0801611290271680  0.4827077690000010
  0.8005533947892687  0.3851185600540390  0.4827077690000010
  0.9987779799346370  0.6145715389356496  0.4827077690000010
  0.1993519065129092  0.6154730282700100  0.4827077690000010
  0.4021378906907174  0.6146192079099251  0.4827077690000010
  0.9957302423235177  0.6913236021973574  0.4827077690000010
  0.5983404867612094  0.6907449536648684  0.4827077690000010
  0.2000610608284390  0.7685103568520049  0.4827077690000010
  0.1986617342543511  0.8455207483716762  0.4827077690000010
  0.7997763520325236  0.8453049167105462  0.4827077690000010
  0.5992014274532806  0.9215648432550410  0.4827077690000010
  0.3998071580927345  0.0019074479358991  0.4827077690000010
  0.6037074319900668  0.0028252803058280  0.4827077690000010
  0.8059632106541486  0.0002105518616844  0.4827077690000010
  0.9976653048257802  0.3056577347195599  0.4827077690000010
  0.5977923774066269  0.3069480787341234  0.4827077690000010
  0.2032218405413033  0.3831916707229297  0.4827077690000010
  0.1972803624545492  0.4610457578627987  0.4827077690000010
  0.5977139158064091  0.4602126125926702  0.4827077690000010
  0.8018884538403128  0.5380117987152886  0.4827077690000010
  0.0013250598179297  0.8467530197854245  0.4827077690000010
  0.5973609911556395  0.8461810799355547  0.4827077690000010
  0.0011875581340774  0.9234557389159121  0.4827077690000010
  0.3966176518946369  0.9209782165065192  0.4827077690000010
  0.0087236607162851  0.0791265746796554  0.4827077690000010
  0.1970629609051286  0.0791980417561007  0.4827077690000010
  0.0050772375592132  0.1554449125644126  0.4827077690000010
  0.5968998554538416  0.2282782679058428  0.4827077690000010
  0.3980581877396148  0.3080514135741481  0.4827077690000010
  0.8003510600383663  0.3068886335286933  0.4827077690000010
  0.9985670219682419  0.3847819666354440  0.4827077690000010
  0.4039106480292247  0.4621263926954455  0.4827077690000010
  0.1940651206432094  0.5369317582882260  0.4827077690000010
  0.4008358570928721  0.5382028122504821  0.4827077690000010
  0.2022449058776630  0.6927072384997359  0.4827077690000010
  0.4007385681392037  0.7691748829628722  0.4827077690000010
  0.5952866882277377  0.7661457837858592  0.4827077690000010
  0.6048818885104978  0.0799656528175845  0.4827077690000010
  0.0000400309793207  0.2304335600819485  0.4827077690000010
  0.3963487715604543  0.2292475422901532  0.4827077690000010
  0.1991150075056041  0.3060294865269526  0.4827077690000010
  0.3980090553487301  0.3829878212193307  0.4827077690000010
  0.5974783027293000  0.3843407649238202  0.4827077690000010
  0.9979926692652157  0.4615881350205058  0.4827077690000010
```

| | | |
|---|---|---|
| 0.6000982620953650 | 0.5398948841820470 | 0.4827077690000010 |
| 0.7967171729331142 | 0.6152247206585031 | 0.4827077690000010 |
| 0.4012168563974114 | 0.6923573232247051 | 0.4827077690000010 |
| 0.7992464459158839 | 0.6921053590569315 | 0.4827077690000010 |
| 0.9946552299571607 | 0.7682984446667760 | 0.4827077690000010 |
| 0.7998658445189193 | 0.7679758776758661 | 0.4827077690000010 |
| 0.1345228429389209 | 0.0288981751389201 | 0.3982269992700864 |
| 0.3343843969164908 | 0.0281171371105771 | 0.3952168673132732 |
| 0.5368461450267077 | 0.0288807445457593 | 0.3959075001293755 |
| 0.7388666833840887 | 0.0272519713100294 | 0.3959609932242927 |
| 0.9405247810435711 | 0.0294900605815869 | 0.3973351932357190 |
| 0.1352118054736238 | 0.1041394946466383 | 0.3987080098221654 |
| 0.3297603319842111 | 0.1042204161673936 | 0.3983915316037709 |
| 0.5390606134310119 | 0.1051740900470008 | 0.3955930952959577 |
| 0.7354688635911657 | 0.1053931214968671 | 0.3954182243151223 |
| 0.9423737140238515 | 0.1051258627934715 | 0.4002861675322222 |
| 0.1333920074735389 | 0.1799247044240602 | 0.3982768360546487 |
| 0.3329633572117956 | 0.1798765480377327 | 0.3955138835036323 |
| 0.5339005871146654 | 0.1801768149115688 | 0.3979310485698804 |
| 0.7337050165266135 | 0.1804680747176803 | 0.3965773440348244 |
| 0.9359490240187824 | 0.1793686165489135 | 0.3990642975161904 |
| 0.1329812445349461 | 0.2555704575124764 | 0.3966845465185926 |
| 0.3355588183201164 | 0.2571528260921809 | 0.3988059896006249 |
| 0.5314720018405055 | 0.2542534645133969 | 0.3984694695991422 |
| 0.7301664691649705 | 0.2560007809637312 | 0.4005155948227497 |
| 0.9303545460678606 | 0.2544378023725997 | 0.3956046956248898 |
| 0.1334952162422880 | 0.3317389029134432 | 0.3954981783966929 |
| 0.3320936139973156 | 0.3317594690659291 | 0.3980678524380608 |
| 0.5286654749680366 | 0.3327953726109598 | 0.3980837448940733 |
| 0.7346106406726207 | 0.3316575881079658 | 0.3976665936514792 |
| 0.9314609288567013 | 0.3335878882642049 | 0.3997455734956148 |
| 0.1306038149503905 | 0.4086545022350023 | 0.3976304904946701 |
| 0.3384605220589449 | 0.4113766403339412 | 0.3975362456246998 |
| 0.5292605237457053 | 0.4084363325557376 | 0.3959957416605562 |
| 0.7293105541437228 | 0.4095303742568532 | 0.3959608998175739 |
| 0.9367367698083697 | 0.4104925974448506 | 0.3972512165911581 |
| 0.1314002332850137 | 0.4877135360186955 | 0.3973813885085775 |
| 0.3368058718178872 | 0.4873886879109079 | 0.4000316239637272 |
| 0.5324692477494466 | 0.4883256542029173 | 0.3980044823679307 |
| 0.7327803741190451 | 0.4860758450400837 | 0.3963389963311670 |
| 0.9311269647128739 | 0.4858097206671346 | 0.3966383791433685 |
| 0.1333562532562169 | 0.5625284291956802 | 0.3978732276420018 |
| 0.3282646484850815 | 0.5610523500675058 | 0.4001919245183174 |
| 0.5306973988070069 | 0.5639328845750597 | 0.3982384660434661 |
| 0.7313255672820560 | 0.5645935978409042 | 0.3967636215940615 |
| 0.9350645351053757 | 0.5626481798130314 | 0.3957582100003947 |
| 0.1347596627462693 | 0.6426609370323320 | 0.3969573538138320 |
| 0.3350495550295349 | 0.6416250999189685 | 0.3960644352972977 |
| 0.5347697534867706 | 0.6396240089518272 | 0.3945568828584385 |
| 0.7332748838008172 | 0.6412898015959669 | 0.3965853658666845 |
| 0.9287543315734865 | 0.6397737027074157 | 0.3954076527626924 |
| 0.1344539336879862 | 0.7162725816756321 | 0.3973059160465340 |
| 0.3334547358659350 | 0.7182314889690815 | 0.4001576342722686 |
| 0.5318053653492640 | 0.7179043847148705 | 0.3969558394302126 |
| 0.7345011231298599 | 0.7176283897094535 | 0.3960478432512318 |
| 0.9301790855165848 | 0.7179232902708677 | 0.3958251594160842 |
| 0.1274124190566468 | 0.7931239009368198 | 0.3961009500964039 |
| 0.3371175735120175 | 0.7938531540411518 | 0.3973686590956689 |
| 0.5328493465700603 | 0.7926706179796952 | 0.3998282338490000 |
| 0.7271372793723785 | 0.7917610854171855 | 0.3986181983714587 |
| 0.9307242000379148 | 0.7939377240078898 | 0.3961749939875077 |
| 0.1342641571253012 | 0.8725344420082379 | 0.3951584038018083 |
| 0.3333542784880379 | 0.8698947654144078 | 0.3951335725951708 |
| 0.5299612042861810 | 0.8706136193398777 | 0.3939175887241717 |
| 0.7340218372959058 | 0.8722891296066209 | 0.3965964967580433 |
| 0.9328348756975444 | 0.8716815921851762 | 0.3944738810656361 |
| 0.1348351949452748 | 0.9492725953384493 | 0.3955847238725454 |
| 0.3329055153234819 | 0.9492667885034294 | 0.3969736729994011 |
| 0.5344526812413903 | 0.9500370463115289 | 0.3968294707683100 |
| 0.7351374691764292 | 0.9489112303729286 | 0.3953505247566937 |
| 0.9345510693856625 | 0.9506759724193472 | 0.3975490813087887 |
| 0.1345228429389209 | 0.0288981751389201 | 0.5671885677299144 |
| 0.3343843969164908 | 0.0281171371105771 | 0.5701986996867277 |
| 0.5368461450267077 | 0.0288807445457593 | 0.5695080668706254 |
| 0.7388666833840887 | 0.0272519713100294 | 0.5694545737757082 |
| 0.9405247810435711 | 0.0294900605815869 | 0.5680803737642819 |
| 0.1352118054736238 | 0.1041394946466383 | 0.5667075571778355 |
| 0.3297603319842111 | 0.1042204161673936 | 0.5670240353962299 |
| 0.5390606134310119 | 0.1051740900470008 | 0.5698224717040432 |
| 0.7354688635911657 | 0.1053931214968671 | 0.5699973426848786 |
| 0.9423737140238515 | 0.1051258627934715 | 0.5651293994677786 |
| 0.1333920074735389 | 0.1799247044240602 | 0.5671387309453522 |
| 0.3329633572117956 | 0.1798765480377327 | 0.5699016834963686 |
| 0.5339005871146654 | 0.1801768149115688 | 0.5674845184301205 |
| 0.7337050165266135 | 0.1804680747176803 | 0.5688382229651765 |
| 0.9359490240187824 | 0.1793686165489135 | 0.5663512694838104 |
| 0.1329812445349461 | 0.2555704575124764 | 0.5687310204814082 |
| 0.3355588183201164 | 0.2571528260921809 | 0.5666095773993760 |
| 0.5314720018405055 | 0.2542534645133969 | 0.5669460974008587 |
| 0.7301664691649705 | 0.2560007809637312 | 0.5648999721772512 |
| 0.9303545460678606 | 0.2544378023725997 | 0.5698108713751111 |
| 0.1334952162422880 | 0.3317389029134432 | 0.5699173886033009 |
| 0.3320936139973156 | 0.3317594690659291 | 0.5673477145619472 |
| 0.5286654749680366 | 0.3327953726109598 | 0.5673318221059276 |
| 0.7346106406726207 | 0.3316575881079658 | 0.5677489733485217 |
| 0.9314609288567013 | 0.3335878882642049 | 0.5656699935043861 |
| 0.1306038149503905 | 0.4086545022350023 | 0.5677850765053236 |
| 0.3384605220589449 | 0.4113766403339412 | 0.5678793213753011 |
| 0.5292605237457053 | 0.4084363325557376 | 0.5694198253394447 |
| 0.7293105541437228 | 0.4095303742568532 | 0.5694546671824270 |
| 0.9367367698083697 | 0.4104925974448506 | 0.5681643504088427 |
| 0.1314002332850137 | 0.4877135360186955 | 0.5680341784914233 |
| 0.3368058718178872 | 0.4873886879109079 | 0.5653839430362737 |
| 0.5324692477494466 | 0.4883256542029173 | 0.5674110846320701 |
| 0.7327803741190451 | 0.4860758450400837 | 0.5690765706688339 |
| 0.9311269647128739 | 0.4858097206671346 | 0.5687771878566323 |
| 0.1333562532562169 | 0.5625284291956802 | 0.5675423393580061 |
| 0.3282646484850815 | 0.5610523500675058 | 0.5652236424816905 |
| 0.5306973988070069 | 0.5639328845750597 | 0.5671771009565347 |
| 0.7313255672820560 | 0.5645935978409042 | 0.5686519454059393 |
| 0.9350645351053757 | 0.5626481798130314 | 0.5696573569996062 |
| 0.1347596627462693 | 0.6426609370323320 | 0.5684582131861617 |
| 0.3350495550295349 | 0.6416250999189685 | 0.5693511317027031 |
| 0.5347697534867706 | 0.6396240089518272 | 0.5708586841415624 |
| 0.7332748838008172 | 0.6412898015959669 | 0.5688302011333235 |
| 0.9287543315734865 | 0.6397737027074157 | 0.5700079142373085 |
| 0.1344539336879862 | 0.7162725816756321 | 0.5681096509534598 |
| 0.3334547358659350 | 0.7182314889690815 | 0.5652579327277323 |
| 0.5318053653492640 | 0.7179043847148705 | 0.5684597275697882 |
| 0.7345011231298599 | 0.7176283897094535 | 0.5693677237487691 |
| 0.9301790855165848 | 0.7179232902708677 | 0.5695904075839167 |
| 0.1274124190566468 | 0.7931239009368198 | 0.5693146169035970 |
| 0.3371175735120175 | 0.7938531540411518 | 0.5680469079043320 |
| 0.5328493465700603 | 0.7926706179796952 | 0.5655873331510008 |
| 0.7271372793723785 | 0.7917610854171855 | 0.5667973686285421 |
| 0.9307242000379148 | 0.7939377240078898 | 0.5692405730124932 |
| 0.1342641571253012 | 0.8725344420082379 | 0.5702571631981925 |
| 0.3333542784880379 | 0.8698947654144078 | 0.5702819944048301 |
| 0.5299612042861810 | 0.8706136193398777 | 0.5714979782758292 |
| 0.7340218372959058 | 0.8722891296066209 | 0.5688190702419575 |
| 0.9328348756975444 | 0.8716815921851762 | 0.5709416859343648 |
| 0.1348351949452748 | 0.9492725953384493 | 0.5698308431274555 |
| 0.3329055153234819 | 0.9492667885034294 | 0.5684418940005997 |
| 0.5344526812413903 | 0.9500370463115289 | 0.5685860962316909 |
| 0.7351374691764292 | 0.9489112303729286 | 0.5700650422433071 |
| 0.9345510693856625 | 0.9506759724193472 | 0.5678664856912121 |

## Supplemental References